\documentclass[aps,prc,superscriptaddress,floats,floatfix,reprint,reprintnumbers]{revtex4}
\usepackage{blindtext}
\usepackage{epsfig}
\usepackage{dcolumn}
\usepackage{bm}
\usepackage{booktabs}
\usepackage{color}
\usepackage{mathrsfs}
\usepackage{ulem}
\usepackage{verbatim}
\usepackage{indentfirst}
\usepackage{float}
\usepackage[colorlinks,
           bookmarks=true,
           linkcolor=blue,
           urlcolor=blue,
            anchorcolor=black,
            citecolor=blue]{hyperref}
\usepackage[bf,raggedright,indentafter]{titlesec}
\usepackage{rotating}
\usepackage{booktabs}
\usepackage{tabularx}
\usepackage{makecell}
\usepackage{hypcap}
\usepackage{tabularx}

\usepackage{array}

\newcommand{\nn}{\nonumber}

\begin{document}
\title{\Large Studying $\mathcal{B}_1(\frac{1}{2}^+)\to \mathcal{B}_2(\frac{1}{2}^+)\ell^+\ell^-$ Semi-leptonic Weak Baryon Decays\\with the SU(3) Flavor Symmetry }
\author{Ru-Min Wang$^{1,\dagger}$,~~Yuan-Guo Xu$^{1,\ddag}$,~~Chong Hua$^{1,\sharp}$,~~Xiao-Dong Cheng$^{2,\S}$\\
$^1${\scriptsize College of Physics and Communication Electronics, JiangXi Normal University, NanChang, JiangXi 330022, China}\\
$^2${\scriptsize College of Physics and Electronic Engineering, XinYang Normal University, XinYang, Henan 464000, China}\\
$^\dagger${\scriptsize ruminwang@sina.com}~~~$^\ddag${\scriptsize yuanguoxv@163.com}~~~$^\sharp${\scriptsize huachongccc@163.com}~~~$^\S${\scriptsize chengxd@mails.ccnu.edu.cn}}

\begin{abstract}
Motivated by recent anomalies in  flavor changing neutral current  $b\to s \ell^+\ell^-$ transitions,  we   study $\mathcal{B}_1\to \mathcal{B}_2\ell^+\ell^-~(\ell=e,\mu,\tau)$  semileptonic weak decays  with the SU(3) flavor symmetry, where  $\mathcal{B}_{1,2}$ are  the spin-$\frac{1}{2}$ baryons of  single bottomed  antitriplet $T_{b3}$,  single charmed antitriplet $T_{c3}$,  or  light baryons octet $T_{8}$.  Using the  SU(3) irreducible representation approach, we first obtain the amplitude relations
among different decay modes  and then    predict the relevant not-yet measured
observables of $T_{b3}\to T_8\ell^+\ell^-$,  $T_{c3}\to T_8\ell^+\ell^-$, and $T_8\to T'_8\ell^+\ell^-$ decays.  {\bf(a)} We calculate the branching ratios of the $T_{b3}\to T_8 \mu^+\mu^-$ and $T_{b3}\to T_8 \tau^+\tau^-$ decay modes  in the whole $q^2$ region and in the different $q^2$ bins by the measurement of $\Lambda^0_b\to \Lambda^0 \mu^+\mu^-$. Many of them  are obtained for the first time. In addition, the longitudinal polarization fractions and the
leptonic forward-backward asymmetries of  all $T_{b3}\to T_{8}\ell^+\ell^-$ decays are very similar to each other in certain $q^2$ bins  due to  the SU(3) flavor symmetry.
{\bf(b)} We analyze the upper limits of  $\mathcal{B}(T_{c3}\to T_{8}\ell^+\ell^-)$  by using the experimental upper limits of  $\mathcal{B}(\Lambda^+_c\to p\mu^+\mu^-)$ and $\mathcal{B}(\Lambda^+_c\to pe^+e^-)$, and find  the experimental upper limit of  $\mathcal{B}(\Lambda^+_c\to p\mu^+\mu^-)$   giving the effective bounds on the relevant SU(3) flavor  symmetry
parameters. The predictions of  $\mathcal{B}(\Xi^0_c \to \Xi^0e^+e^-)$ and  $\mathcal{B}(\Xi^0_c \to \Xi^0\mu^+\mu^-)$  will be different between  the single-quark transition dominant contributions   and  the $W$-exchange dominant ones.
{\bf(c)} As for $T_{8}\to T'_8 \ell^+\ell^-$  decays,  we analyze the  single-quark transition  contributions   and  the $W$-exchange   contributions by  using  the two experimental measurements of  $\mathcal{B}(\Xi^0\to \Lambda^0 e^+e^-)$ and $\mathcal{B}(\Sigma^+\to p\mu^+\mu^-)$, and give the branching ratio predictions by assuming  either single-quark transition dominant contributions  or the $W$-exchange dominant contributions.
According to our predictions,  some observables  are accessible to the experiments at BESIII, LHCb and Belle-II.

\end{abstract}

\maketitle

\section{INTRODUCTION}
Flavor changing neutral current (FCNC) processes,  such as $b\to s \ell^+\ell^-$,    give access to important tests of  the standard model (SM) and searches  for  new physics  beyond the SM.
Recently, some discrepancies with the SM are reported in several observables in  $B$ meson decays, for example,  the angular-distribution
observable $P'_5$ of $B^0\to K^{*0}\mu^+\mu^-$ \cite{DescotesGenon:2012zf,Aaij:2015oid,ATLAS:2017dlm,Sirunyan:2017dhj} and the lepton flavor universality observables $R_{K^+}$ and $R_{K^{*0}}$ with $R_M=\frac{\mathcal{B}(B\to M\mu^+\mu^-)}{\mathcal{B}(B\to Me^+e^-)}$ \cite{Aaij:2014ora,Aaij:2017vbb}.
Semileptonic baryon  decays are quite  different to $B,D,K$ meson ones, for instance,  the initial baryons
may be polarized,  the transitions involve a diquark system as a spectator
rather than a single-quark, and the $W$-exchange contributions
of two-quark and three-quark transitions might appear in baryon decays. Therefore, the baryon
decays   provide the important additional tests of the SM predictions, which can be used to improve the understanding of recent anomalies  in
  $B$ meson decays.  Recently significant experimental progress has been achieved in studying   rare $\Lambda_b$ decays.
The $\Lambda_b\to \Lambda\mu^+\mu^-$ baryon decay is  the only one measured   among  the $T_{b3}\to T_8\ell^+\ell^-$ decays  at present.  $\mathcal{B}(\Lambda_b\to \Lambda\mu^+\mu^-)$ was first measured by the CDF Collaboration \cite{Aaltonen:2011qs} and then greatly  improved by LHCb   \cite{Aaij:2015xza,Aaij:2013hna}. For $T_{c3}\to T_8\ell^+\ell^-$ decays, only $\mathcal{B}(\Lambda_c\to pe^+e^-)$  and $\mathcal{B}(\Lambda_c\to p\mu^+\mu^-)$ have been upper limited by $BABAR$ and  LHCb \cite{Lees:2011hb,Aaij:2017nsd}.   As for $T_8\to T'_8\ell^+\ell^-$ decays,  $\Xi^0\to \Lambda^0 e^+e^-$ and $\Sigma^+\to p\mu^+\mu^-$ have been measured by  NA48
 \cite{Batley:2007hp} and  HyperCP  \cite{Park:2005eka}, respectively.
With the experiment development,
some $\mathcal{B}_1\to \mathcal{B}_2\ell^+\ell^-$  decays will  be improved or detected by the BESIII, LHCb, and Belle-II Collaborations in the near future, so
it is necessary to study $\mathcal{B}_1\to \mathcal{B}_2\ell^+\ell^-$ decays   theoretically.

The theoretical challenge in the study of $\mathcal{B}_1\to \mathcal{B}_2\ell^+\ell^-$ decays is  calculating the hadronic  $\mathcal{B}_1 \to \mathcal{B}_2 $
form factors in the hadronic matrix elements.
Form factors for $\Lambda_b\to \Lambda$ have been estimated in lattice QCD \cite{Detmold:2016pkz,Boer:2014kda,Gutsche:2013pp}, QCD light cone sum rules \cite{Wang:2015ndk}, the soft-collinear effective theory \cite{Feldmann:2011xf}, and perturbative QCD \cite{He:2006ud}. Form factors for $\Lambda_b\to n$ have been estimated in the relativistic quark diquark picture  \cite{Faustov:2017ous} and the context of light
cone QCD sum rules \cite{Azizi:2010zzb,Aliev:2010uy}.   Nevertheless, other form factors of  $T_{b3}\to T_8\ell^+\ell^-$, such as the ones for  $\Xi^0_b\to \Xi^0$, $\Xi^-_b\to \Xi^-$,  $\Xi^0_b\to \Lambda^0$, $\Xi^0_b\to \Sigma^0$ and $\Xi^-_b\to \Sigma^-$, have not been calculated  yet.  Similarly  in $T_{c3}\to T_8\ell^+\ell^-$ and $T_8\to T'_8\ell^+\ell^-$ decays,  only some form factors are calculated, for example,  ones for $\Lambda^+_c\to p$ transition \cite{Azizi:2010zzb,Sirvanli:2016wnr,Meinel:2017ggx,Faustov:2018dkn}.

Theoretical
calculations of the hadronic matrix elements are not well understood due to our poor understanding of QCD at low energy regions.
 The SU(3) flavor  symmetry approach is independent of the detailed dynamics offering us  an opportunity to relate different decay modes.
Nevertheless, it cannot determine
the sizes of the amplitudes by itself. However, if experimental data are enough, one may use the data to extract the  amplitudes, which can be viewed as predictions based on symmetry. Although
 SU(3) flavor symmetry is only an approximate symmetry because $u$,
$d$ and $s$ quarks have different masses, it still provides some useful information
about the decays.
One popular way of predicting the SU(3) flavor symmetry is to construct the
SU(3) irreducible representation amplitude by decomposing the effective Hamiltonian, in which  one only focuses on the SU(3)
flavor structure of the initial states and finial states, but does not involve the details about the
interaction dynamics.

Some $\mathcal{B}_1(\frac{1}{2}^+)\to \mathcal{B}_2(\frac{1}{2}^+)\ell^+\ell^-$ semileptonic baryon decays have been well studied, for instance,  semileptonic $\Lambda^0_b$ decays in Refs. \cite{Detmold:2016pkz,Boer:2014kda,Gutsche:2013pp,Wang:2008ni,Hu:2017qxj,Roy:2017dum,Das:2018iap,Aslam:2008hp,Faustov:2017wbh},  semileptonic $\Lambda^+_c$ decays in Refs. \cite{deBoer:2017way,Faustov:2018dkn,Sirvanli:2016wnr}, and semileptonic  $\Sigma^+$ decays in Refs. \cite{He:2006fr,He:2005yn,He:2018yzu,Xiangdong:2007vv,Bergstrom:1987wr}. In this work, we will study all weak $\mathcal{B}_1(\frac{1}{2}^+)\to \mathcal{B}_2(\frac{1}{2}^+)\ell^+\ell^-$ decays by using the  SU(3) irreducible representation approach. We first obtain the amplitude relations
among different decay modes, then use the available data to extract the SU(3) irreducible amplitudes   and finally  predict the  not-yet-measured modes for further tests in experiments.

This paper is organized as follows. In Sec. II, we will collect the representations for the baryon multiplets of $\frac{1}{2}$-spin and the observable expressions  of relevant baryon decays. In Sec. III, we will analyze the semileptonic weak  decays of $T_{b3}\to T_8\ell^+\ell^-$, $T_{c3}\to T_8\ell^+\ell^-$, and $T_8\to T'_8\ell^+\ell^-$.   Our conclusions are given in Sec. IV.

\section{Theoretical Frame}

\subsection{Baryon multiplets with $\frac{1}{2}$ spin}
The light baryons octet $T_{8}$ under the SU(3) flavor symmetry of $u,d,s$ quarks  can be written as
\begin{eqnarray}
 T_8&=&\left(\begin{array}{ccc}
\frac{\Lambda^0}{\sqrt{6}}+\frac{\Sigma^0}{\sqrt{2}} & \Sigma^+ & p \\
\Sigma^- &\frac{\Lambda^0}{\sqrt{6}}-\frac{\Sigma^0}{\sqrt{2}}  & n \\
\Xi^- & \Xi^0 &-\frac{2\Lambda^0}{\sqrt{6}}
\end{array}\right)\,.
\end{eqnarray}
The single charmed antitriplet $T_{c3}$  is given as
\begin{eqnarray}
T_{c3}&=&(\Xi^0_c,~-\Xi^+_c,~\Lambda^+_c).
\end{eqnarray}
The  antitriplet $T_{b3}$  with a heavy b quark is
\begin{eqnarray}
T_{b3}&=&(\Xi^-_b,~-\Xi^0_b,~\Lambda^0_b).
\end{eqnarray}

\subsection{Helicity amplitudes for semileptonic decays }
 In the SM, the low energy effective Hamiltonians for  $b\to s/d\ell^+\ell^-$, $c\to u\ell^+\ell^-$ and  $s\to d\ell^+\ell^-$ FCNC transitions have similar forms and can be written as  \cite{Buchalla:1995vs,Meinel:2017ggx,deBoer:2015boa,Shifman:1976de,He:2005yn}
\begin{eqnarray}
\mathcal{H}(q_1\to q_2\ell^+\ell^-)=-\frac{\alpha_eG_F}{\sqrt{2}\pi}\lambda_{q_1q_2}\left(C^{eff}_9\bar{q}_2\gamma^\mu P_L q_1 \bar{\ell}\gamma_\mu\ell +C_{10}\bar{q}_2\gamma^\mu P_Lq_1 \bar{\ell}\gamma_\mu\gamma_5\ell-\frac{2m_{q_1}}{q^2}C^{eff}_7 \bar{q}_2iq_\nu\sigma^{\mu\nu}P_R q_1 \bar{\ell}\gamma_\mu\ell\right),\label{EQ:Heff}
\end{eqnarray}
where $G_F$ denotes the Fermi constant, the fine structure constant $\alpha_e=\frac{e^2}{4\pi}$, the chiral projection operators $P_{L,R}=(1\mp\gamma_5)/2$, $\sigma_{\mu\nu}=\frac{i[\gamma_\mu,\gamma_\nu]}{2}$, $\lambda_{q_1q_2}$  denotes the Cabibbo-Kobayashi-Maskawa (CKM) elements, and $C_{i}$  denote Wilson coefficients.
For the $b\to s/d\ell^+\ell^-$ transitions via the  $u\bar{u},c\bar{c}$ loops, $\lambda_{bs(bd)}=V_{tb}V^*_{ts}$($V_{tb}V^*_{td}$), and the expressions of Wilson coefficients $C^{eff}_{7,9}$ and $C_{10}$  are given in Ref. \cite{Buchalla:1995vs}.  For the $c\to u\ell^+\ell^-$ transition via the $d\bar{d},s\bar{s}$ loops, $\lambda_{cu}C^{eff}_{7,9}=\frac{4\pi}{\alpha_s}\Big[V^*_{cd}V_{ud}C^{eff(d)}_{7,9}(q^2)+V^*_{cs}V_{us}C^{eff(s)}_{7,9}(q^2)\Big]$ as well as  $C_{10}=0$  and the  expressions of  $C^{eff(s,d)}_{7,9}(q^2)$ can be found in  Refs. \cite{Meinel:2017ggx,deBoer:2015boa}.  For the $s\to d\ell^+\ell^-$ transition via the $u\bar{u}$ loop, $\lambda_{sd}=V_{us}V^*_{ud}$, $C^{eff}_7\approx \frac{V_{cs}V^*_{cd}}{2V_{us}V^*_{ud}}c^c_{7\gamma}$,  $C^{eff}_9=(z_{7V}-\frac{V_{ts}V^*_{td}}{V_{us}V^*_{ud}} y_{7V})\frac{2\pi}{\alpha_e}$, and   $C_{10}=-\frac{V_{ts}V^*_{td}}{V_{us}V^*_{ud}}\frac{2\pi}{\alpha_e}y_{7A}$ with $z_{7V}=-0.046\alpha_e$, $y_{7V}=0.735\alpha_e$, $y_{7A}=-0.700\alpha_e$ as well as  $c^c_{7\gamma}=0.13\alpha_e$ from Refs. \cite{Buchalla:1995vs,Shifman:1976de,He:2005yn}.
 For  $T_{b3}\to T_8\ell^+\ell^-$ and  $T_{c3}\to T_8\ell^+\ell^-$ decays, the Wilson coefficient $C^{eff}_9$ receives not only  from the four quark operators but also from the long distance (LD) contributions coming from $c\bar{c}$ for $T_{b3}\to T_8\ell^+\ell^-$ and $d\bar{d},s\bar{s}$ for $T_{c3}\to T_8\ell^+\ell^-$.  Note that the $T_{c3}\to T_8\ell^+\ell^-$ decays are
dominated by LD contributions.  For $T_{8}\to T'_8\ell^+\ell^-$ decays, the LD contribution arises mainly from the photon-mediated process $T_8\to T'_8\gamma^*\to T'_8\ell^+\ell^-$ \cite{He:2018yzu,Bergstrom:1987wr}.

The helicity  amplitudes for $\mathcal{B}_1\to \mathcal{B}_2\ell^+\ell^-$   can be obtained  from  Eq. (\ref{EQ:Heff}), 
\begin{equation}
\mathcal{M}^{\lambda_1,\lambda_2}_{VA}(s_p,s_k)=-\frac{G_F}{\sqrt{2}}\frac{\alpha_e}{4\pi}\lambda_{q_1q_2}\sum_\lambda \eta_\lambda \left[H^{L,s_p,s_k}_{ VA,\lambda}L^{\lambda_1,\lambda_2}_{L,\lambda}+H^{R,s_p,s_k}_{ VA,\lambda}L^{\lambda_1,\lambda_2}_{R,\lambda} \right],
\end{equation}
with
\begin{eqnarray}
L^{\lambda_1\lambda_2}_{L(R),\lambda}&=&\bar{\epsilon}^\mu(\lambda)\left\langle\bar{\ell}(\lambda_1)\ell(\lambda_2)\left|\bar{\ell}\gamma_\mu(1\mp\gamma_5)\ell\right|0\right\rangle,\nonumber\\
H^{L(R),s_p,s_k}_{VA,\lambda}&=&\bar{\epsilon}^*_\mu(\lambda)\left\langle \mathcal{B}_2(k,s_k)\left|\left[\Big(C^{eff}_9\mp C_{10}\Big)\bar{q}_2\gamma^\mu(1-\gamma_5)q_1-\frac{2m_{q_1}}{q^2}C^{eff}_7 \bar{q}_2iq_\nu\sigma^{\mu\nu}(1+\gamma_5)q_1\right] \right| \mathcal{B}_1(p,s_p)\right\rangle,
\end{eqnarray}
where $q^2\equiv (p-k)^2$ bounded in physical region as $ (2m_\ell)^2\leq q^2\leq (m_{\mathcal{B}_1}-m_{\mathcal{B}_2})^2$, the polarization of the gauge boson $\lambda=t,\pm1,0$, the helicities of the final state leptons are $\lambda_{1,2}$, and $\eta_t=1,\eta_{\pm1,0}=-1$.

The nonvanishing leptonic helicity amplitudes $L^{\lambda_1\lambda_2}_{L(R),\lambda}$ are
\begin{eqnarray}
L^{-\frac{1}{2}+\frac{1}{2}}_{ L,+1}&=&-L^{+\frac{1}{2}-\frac{1}{2}}_{ R,-1} =\sqrt{\frac{q^2}{2}} (1+\beta_\ell)(1+cos\theta_\ell),\nn\\
L^{+\frac{1}{2}-\frac{1}{2}}_{ R,+1}&=&-L^{-\frac{1}{2}+\frac{1}{2}}_{ L,-1} =-\sqrt{\frac{q^2}{2}} (1+\beta_\ell)(1-cos\theta_\ell),\nn\\
L^{-\frac{1}{2}+\frac{1}{2}}_{ L,0}&=&L^{+\frac{1}{2}-\frac{1}{2}}_{ R,0} =\sqrt{q^2} (1+\beta_\ell)sin\theta_\ell,
\end{eqnarray}
with $\beta_\ell=\sqrt{1-\frac{4m^2_\ell}{q^2}}$.

The $\mathcal{B}_1 \to \mathcal{B}_2$ hadronic matrix elements are calculated in the frameworks
of soft-collinear effective theory \cite{Feldmann:2011xf} and lattice QCD \cite{Detmold:2016pkz,Boer:2014kda,Gutsche:2013pp}. The helicity-based definition of the form factors are  presented as \cite{Detmold:2016pkz,Boer:2014kda,Gutsche:2013pp}
{\small
\begin{eqnarray}
\langle \mathcal{B}_2(k,s_k)|\bar{q}_2\gamma^\mu q_1 |\mathcal{B}_1\rangle &=& \bar{u}(k,s_k)\Bigg[f_0(q^2)(m_{B_1}-m_{B_2})\frac{q^\mu}{q^2}\nonumber + f_+(q^2) \frac{m_{B_1}+m_{B_2}}{s_+} \left\{p^\mu + k^\mu  - \frac{q^\mu}{q^2}(m_{B_1} - m_{B_2}) \right\} \nonumber\\
&& ~~~~~~~~~~~+  f_\perp(q^2) \left\{ \gamma^\mu - \frac{2m_{B_2}}{s_+}p^\mu - \frac{2m_{B_1}}{s_+}k^\mu \right\} \Bigg]u(p,s_p)\, ,\\
\langle \mathcal{B}_2(k,s_k)|\bar{q}_2\gamma^\mu\gamma_5 q_1 |\mathcal{B}_1(p,s_p)\rangle &=& - \bar{u}(k,s_k) \gamma_5 \Bigg[ g_0(q^2) (m_{B_1} + m_{B_2}) \frac{q^\mu}{q^2} + g_+(q^2) \frac{m_{B_1} - m_{B_2}}{s_-} \left\{p^\mu + k^\mu - \frac{q^\mu}{q^2} (m_{B_1} - m_{B_2}) \right\} \nonumber\\
&&~~~~~~~~~~~~~~~~ + g_\perp(q^2) \left\{\gamma^\mu + \frac{2m_{B_2}}{s_-}p^\mu - \frac{2m_{B_1}}{s_-}k^\mu \right\}  \Bigg] u(p,s_p)\, ,\\
\langle\mathcal{B}_2(p',s')| \bar q_2 \, i\sigma_{\mu\nu} q^\nu \, q_1|\mathcal{B}_1(p,s)\rangle
&=& - \bar{u}_{\mathcal{B}_2}(p',s')
 \left\{
 h_+(q^2)\frac{q^2}{s_+}
\left( p_\mu + p_\mu' - \frac{q_\mu}{q^2} (m_{\mathcal{B}_1}^2-m_{\mathcal{B}_2}^2) \right)\right.\nonumber\\
&&\left.+(m_{\mathcal{B}_1}+m_{\mathcal{B}_2}) h_\perp(q^2)
\left( \gamma_\mu -  \frac{2  m_{\mathcal{B}_2}}{s_+}  p_\mu - \frac{2m_{\mathcal{B}_1}}{s_+} p'_\mu
  \right)
 \right\} u_{\mathcal{B}_1}(p,s) ,\\
\langle \mathcal{B}_2(p',s')| \bar q_2 \, i\sigma_{\mu\nu} \gamma_5 q^\nu q_1|\mathcal{B}_1\rangle
&=& -  \bar u_{\mathcal{B}_2}(p',s')
 \gamma_5 \left\{
 \tilde h_+(q^2)  \frac{q^2}{s_-}
\left( p_\mu + p_\mu' - \frac{q_\mu}{q^2}  (m_{\mathcal{B}_1}^2-m_{\mathcal{B}_2}^2) \right)\right.\nonumber\\
&&\left.
+ (m_{\mathcal{B}_1}-m_{\mathcal{B}_2}) \tilde h_\perp(q^2) \left( \gamma_\mu +
 \frac{2 m_{\mathcal{B}_2}}{s_-}  p_\mu - \frac{2 m_{\mathcal{B}_1}}{s_-} p'_\mu
  \right)
 \right\}  u_{\mathcal{B}_1}(p,s),
\end{eqnarray}}
where $s_\pm = (m_{\mathcal{B}_1}\pm m_{\mathcal{B}_2})^2-q^2$ and $f^{V,A,T,T_5}_{0,\perp}$  are the form factors. And then we obtain the nonvanishing hadronic helicity amplitudes $H^{L(R),s_p,s_k}_{VA,\lambda}$
{\small
\begin{eqnarray}
H^{L(R),+\frac{1}{2}+\frac{1}{2}}_{ VA,0} &=& f_+(q^2)(m_{\mathcal{B}_1} + m_{\mathcal{B}_2}) \sqrt{\frac{s_-}{q^2}} \mathcal{C}^{L(R)}_{ VA} - g_+(q^2)(m_{\mathcal{B}_1} - m_{\mathcal{B}_2}) \sqrt{\frac{s_+}{q^2}} \mathcal{C}^{L(R)}_{ VA} + \frac{2m_b}{q^2} \bigg( h_+(q^2) \sqrt{q^2 s_-} - \tilde{h}_+(q^2) \sqrt{q^2 s_+} \bigg)  \mathcal{C}_7^{ eff}\, ,\nonumber\\
H^{L(R),-\frac{1}{2}-\frac{1}{2}}_{ VA,0} &=& f_+(q^2)(m_{\mathcal{B}_1} + m_{\mathcal{B}_2}) \sqrt{\frac{s_-}{q^2}} \mathcal{C}^{L(R)}_{ VA} + g_+(q^2)(m_{\mathcal{B}_1} - m_{\mathcal{B}_2}) \sqrt{\frac{s_+}{q^2}} \mathcal{C}^{L(R)}_{ VA} + \frac{2m_b}{q^2} \bigg( h_+(q^2) \sqrt{q^2 s_-} + \tilde{h}_+(q^2) \sqrt{q^2 s_+} \bigg)  \mathcal{C}_7^{ eff}\, ,\nonumber\\
H^{L(R),-\frac{1}{2}+\frac{1}{2}}_{ VA,+} &=& -f_\perp(q^2) \sqrt{2s_-} \mathcal{C}^{L(R)}_{ VA} + g_\perp(q^2) \sqrt{2s_+} \mathcal{C}^{L(R)}_{ VA} - \frac{2m_b}{q^2} \bigg( h_\perp(q^2)(m_{\mathcal{B}_1} + m_{\mathcal{B}_2}) \sqrt{2s_-} - \tilde{h}_\perp(q^2) (m_{\mathcal{B}_1} - m_{\mathcal{B}_2}) \sqrt{2s_+}   \bigg)\mathcal{C}_7^{ eff}\,,\nonumber\\
H^{L(R),+\frac{1}{2}-\frac{1}{2}}_{ VA,-} &=& -f_\perp(q^2) \sqrt{2s_-} \mathcal{C}^{L(R)}_{ VA} - g_\perp(q^2) \sqrt{2s_+} \mathcal{C}^{L(R)}_{ VA}- \frac{2m_b}{q^2} \bigg( h_\perp(q^2)(m_{\mathcal{B}_1} + m_{\mathcal{B}_2}) \sqrt{2s_-} + \tilde{h}_\perp(q^2) (m_{\mathcal{B}_1} - m_{\mathcal{B}_2}) \sqrt{2s_+}   \bigg)\mathcal{C}_7^{ eff}\,,\nonumber\\\label{EQ:H}
\end{eqnarray}}
with $\mathcal{C}^{L(R)}_{ VA}\equiv (\mathcal{C}_9^{ eff}\mp \mathcal{C}_{10})$.

In addition, in terms  of the  SU(3) flavor symmetry, baryon states and quark operators can be parametrized into SU(3) tensor forms, while the polarization vectors $\bar{\epsilon}^*(\lambda)$ and leptonic helicity amplitudes $L^{\lambda_1,\lambda_2}_{L(R),\lambda}$ are invariant under SU(3)  flavor symmetry. The hadronic helicity amplitude relations of $\mathcal{B}_1\to \mathcal{B}_2\ell^+\ell^-$ are similar to ones of  $\mathcal{B}_1\to \mathcal{B}_2\gamma$ as given in Ref. \cite{Wang:2020wxn}, and  will be given in next section for convenience.

\subsection{Observables for  $\mathcal{B}_1\to \mathcal{B}_2\ell^+\ell^-$}

In
the rest frame of the baryon $\mathcal{B}_1$, the double differential decay branching ratio is \cite{Das:2018sms}
\begin{eqnarray}
\frac{d\mathcal{B}(\mathcal{B}_1\to \mathcal{B}_2\ell^+\ell^-)}{dq^2 d\cos\theta_\ell} &=& \frac{\tau_{\mathcal{B}_1}}{2m^3_{\Lambda_b}} \frac{2\sqrt{\lambda(m^2_{\mathcal{B}_1},m^2_{\mathcal{B}_2},q^2)}}{(8\pi)^3}\frac{1}{2s_p+1}\sum_{\lambda_1,\lambda_2}\sum_{s_p,s_k}\big|\mathcal{M}^{\lambda_1,\lambda_2}_{VA}(s_p,s_k)\big|^2\, \nonumber\\
&=& N^2(q^2)  \bigg[ (1 - cos^2\theta_\ell) \bigg( \big|H_{ VA,0}^{L,+\frac{1}{2}+\frac{1}{2}}\big|^2 + \big|H_{ VA,0}^{L,-\frac{1}{2}-\frac{1}{2}}\big|^2 + \big|H_{ VA,0}^{R,+\frac{1}{2}+\frac{1}{2}}\big|^2 + \big|H_{ VA,0}^{R,-\frac{1}{2}-\frac{1}{2}}\big|^2\bigg)\nonumber\\
&& + \frac{1}{2}(1-cos\theta_\ell)^2\bigg( \big| H_{ VA,-}^{L,+\frac{1}{2}-\frac{1}{2}} \big|^2 + \big| H_{ VA,+}^{R,-\frac{1}{2}+\frac{1}{2}} \big|^2\bigg) \nonumber\\
&& +\frac{1}{2}(1+cos\theta_\ell)^2 \bigg( \big| H_{ VA,-}^{R,+\frac{1}{2}-\frac{1}{2}} \big|^2 + \big| H_{ VA,+}^{L,-\frac{1}{2}+\frac{1}{2}} \big|^2 \bigg) \bigg],\label{EQ:ddB}
\end{eqnarray}
with
\begin{equation}
N(q^2) = G_F \lambda_{q_1q_2} \alpha_e \left(1+\sqrt{1-\frac{4m^2_\ell}{q^2}}\right) \sqrt{\tau_{\mathcal{B}_1} \frac{q^2\sqrt{\lambda(m_{\mathcal{B}_1},m_{\mathcal{B}_2},q^2)}}{2^{15} m^3_{\mathcal{B}_1} \pi^5 }  },
\end{equation}
where $\lambda(a,b,c)=a^2+b^2+c^2-2(ab+bc+ca)$.
And the  differential decay branching ratio is
\begin{eqnarray}
\frac{d\mathcal{B}(\mathcal{B}_1\to \mathcal{B}_2\ell^+\ell^-)}{dq^2}= \frac{4}{3}N^2(q^2) H_M(q^2),\label{EQ:dB}
\end{eqnarray}
with
\begin{eqnarray}
H_M(q^2)&=&\bigg( \big| H_{ VA,-}^{L,+\frac{1}{2}-\frac{1}{2}} \big|^2 + \big| H_{ VA,+}^{R,-\frac{1}{2}+\frac{1}{2}} \big|^2\bigg)+\bigg( \big| H_{ VA,-}^{R,+\frac{1}{2}-\frac{1}{2}} \big|^2 + \big| H_{ VA,+}^{L,-\frac{1}{2}+\frac{1}{2}} \big|^2 \bigg)\nonumber\\
&&+\bigg( \big|H_{ VA,0}^{L,+\frac{1}{2}+\frac{1}{2}}\big|^2 + \big|H_{ VA,0}^{L,-\frac{1}{2}-\frac{1}{2}}\big|^2 + \big|H_{ VA,0}^{R,+\frac{1}{2}+\frac{1}{2}}\big|^2 + \big|H_{ VA,0}^{R,-\frac{1}{2}-\frac{1}{2}}\big|^2\bigg).\label{EQ:HMq2}
\end{eqnarray}

The longitudinal polarization fraction can be obtained by Eq. (\ref{EQ:ddB})
\begin{equation}
F_L(q^2) = \frac{ \displaystyle\int_{-1}^{+1} d\cos\theta_\ell (2-5\cos^2\theta_\ell) \frac{d^2\mathcal{B}}{dq^2d\cos\theta_\ell} }{ \displaystyle\int_{-1}^{+1} d\cos\theta_\ell \frac{d^2\mathcal{B}}{dq^2d\cos\theta_\ell} }\, .
\end{equation}
and the concrete expression is
\begin{eqnarray}
F_L(q^2) &=&\bigg( \big|H_{ VA,0}^{L,+\frac{1}{2}+\frac{1}{2}}\big|^2 + \big|H_{ VA,0}^{L,-\frac{1}{2}-\frac{1}{2}}\big|^2 + \big|H_{ VA,0}^{R,+\frac{1}{2}+\frac{1}{2}}\big|^2 + \big|H_{ VA,0}^{R,-\frac{1}{2}-\frac{1}{2}}\big|^2\bigg)
\left[H_M(q^2)\right]^{-1}.\label{Eq:fLc}
\end{eqnarray}

The leptonic forward-backward asymmetry
\begin{equation}
A^\ell_{ FB}(q^2) = \frac{ \displaystyle\int_{0}^{+1} d\cos\theta_\ell \frac{d^2\mathcal{B}}{dq^2d\cos\theta_\ell} - \displaystyle\int_{-1}^{0} d\cos\theta_\ell \frac{d^2\mathcal{B}}{dq^2d\cos\theta_\ell} }{ \displaystyle\int_{-1}^{0} d\cos\theta_\ell \frac{d^2\mathcal{B}}{dq^2d\cos\theta_\ell} + \displaystyle\int_{0}^{+1} d\cos\theta_\ell \frac{d^2\mathcal{B}}{dq^2d\cos\theta_\ell} }\, ,
\end{equation}
and the concrete expression is
\begin{eqnarray}
A^\ell_{ FB}(q^2) =&& \frac{3}{4}\left[\bigg( \big| H_{ VA,-}^{R,+\frac{1}{2}-\frac{1}{2}} \big|^2 + \big| H_{ VA,+}^{L,-\frac{1}{2}+\frac{1}{2}} \big|^2 \bigg)-\bigg( \big| H_{ VA,-}^{L,+\frac{1}{2}-\frac{1}{2}} \big|^2 + \big| H_{ VA,+}^{R,-\frac{1}{2}+\frac{1}{2}} \big|^2\bigg)\right]
\left[H_M(q^2)\right]^{-1}.\label{Eq:AlFBc}
\end{eqnarray}

The lepton flavor universality in baryon weak decays $T_{b3}\to T_8\ell^+\ell^-$  is defined in
a manner identical $R_{K^{(*)}}$ as
\begin{eqnarray}
R_{T_{b3}\to T_8}\equiv\frac{\int^{q_{max}}_{q_{min}}d\mathcal{B}(T_{b3}\to T_8\mu^+\mu^-)/ds}{\int^{q_{max}}_{q_{min}}d\mathcal{B}(T_{b3}\to T_8e^+e^-)/ds}.
\end{eqnarray}

For  $q^2$ integration of $X(q^2)=F_L(q^2)$ and $A^\ell_{ FB}(q^2)$, following Ref. \cite{Bobeth:2010wg}, two ways of integration are considered.
The normalized $q^2$-integrated observables $\langle X \rangle$  are calculated by separately
integrating the numerators and denominators with the same $q^2$ bins.
The ``naively integrated" observables are obtained by
\begin{eqnarray}
\overline{X} =\frac{1}{q^2_{max}-q^2_{min}}\int^{q^2_{max}}_{q^2_{min}}dq^2X(q^2).
\end{eqnarray}

Note that, besides the single-quark transition contributions, the $W$-exchange contributions via the two-quark and three-quark transitions as well as the internal radiation transition, which contribute to the radiative baryon decays $\mathcal{B}_1\to \mathcal{B}_2\gamma$ \cite{Verma:1988gf,Azimov:1996uf,Lach:1995we}, may also contribute to the semileptonic baryon decays $\mathcal{B}_1 \to \mathcal{B}_2\ell^+\ell^-$. In some decays, for example, $\Sigma^+\to p \ell^+\ell^-$ decays, the $W$-exchange contributions with the two-quark transition will play a major role \cite{Bergstrom:1987wr}. So we will consider these W-exchange contributions   in the later analysis of SU(3) flavor symmetry.

\section{Results and Analysis}
The theoretical input parameters and the experimental
data within the $1\sigma$ error from the Particle Data Group \cite{PDG2020} will
be used in our numerical results.
To obtain SU(3) IRA amplitudes, one just needs to contract all upper and lower indices of the hadrons
and the Hamiltonian to form all possible SU(3) singlets and associate each  with a parameter which lumps up the
Wilson coefficients and unknown hadronization effects \cite{Pan:2020qqo}. These parameters can be determined theoretically and
experimentally. In this work, we will determine these parameters by relevant experimental data, and then give the predictions for other not-yet-measured decay modes.
For $T_{b3 }$ semileptonic decays, there are enough phase spaces to allow for $e^+e^-,\mu^+\mu^-$, and $\tau^+\tau^-$
decays.  $T_{c3 }$ and $T_{8}$  semileptonic decays only have
enough phase spaces to allow for both $e^+e^-$ and $\mu^+\mu^-$
decays.
The results of $T_{b3}\to T_8e^+e^-,T_8\mu^+\mu^-,T_8\tau^+\tau^-$ decays, $T_{c3}\to T_8e^+e^-,T_8\mu^+\mu^-$ decays, and  $T_8\to T'_8e^+e^-,T'_8\mu^+\mu^-$ decays are given in the following subsections A,  B, and  C, respectively.

\subsection{$T_{b3}\to T_8\ell^+\ell^-$   weak decays}\label{Sec:Tb32T8ll}
For  $T_{b3}\to T_8\ell^+\ell^-$ decays, the single-quark transition contributions are  strongly dominant so
that other contributions like the $W$-exchange contributions are usually omitted.
The SU(3) flavor structure  Hamiltonian with $b\to s,d$  transitions can been found, for instance,  in Refs. \cite{Zeppenfeld:1980ex,Savage:1989ub,Deshpande44}, and  the  SU(3) IRA  hadronic helicity amplitudes for $T_{b3}\to T_{8}\ell^+\ell^-$ via $b\to s/d \ell^+\ell^-$ can be parametrized as
\begin{eqnarray}
H(T_{b3} \to T_8\ell^+\ell^-)^{L(R),s_p,s_k}_{VA,\lambda}&=&e_1(T_{b3})^{[ij]}T(\bar{3})^k(T_8)_{[ij]k}+e_2(T_{b3})^{[ij]}T(3)^k(T_8)_{[ik]j},\label{Eq:HTb32T8}
\end{eqnarray}
which are similar to the decay amplitudes of corresponding $T_{b3}\to T_{8}\gamma$ modes in Ref. \cite{Wang:2019alu}. In Eq. (\ref{Eq:HTb32T8}), $T(\bar{3})=(0,1,1)$  denoted the transition operators $(\bar{q}_2b)$  with $q_2=s,d$, and the model as well as scale independent parameters $e_i\equiv(e_i)^{L(R),s_p,s_k}_{VA,\lambda}(q^2)$. The parameters $e_i$ contain information about QCD dynamics,  and could include the long distance (LD) contributions from hadron resonances.  The SU(3) IRA amplitudes of the  $T_{b3}\to T_{8}\ell^+\ell^- $ weak decays are given in Tab. \ref{Tab:SU3HATb3}, and for a better understanding, the information of
relevant CKM matrix elements  are also listed in Tab. \ref{Tab:SU3HATb3}.
From  Tab. \ref{Tab:SU3HATb3}, one can see that $\Lambda^0_b\to \Sigma^0\ell^+\ell^-$ decays are not allowed by the  SU(3) flavor symmetry, and other decay modes via $b\to s/d\ell^+\ell^-$ can be related by only  one parameter $E\equiv e_1+e_2$.

\begin{table}[t]
\renewcommand\arraystretch{1.3}
\tabcolsep 0.25in
\centering
\caption{The SU(3) IRA amplitudes of the  $T_{b3}\to T_{8}\ell^+\ell^- $ weak decays by the $b\to s/d \ell^+\ell^-$ transitions, and $E\equiv e_1+e_2$. }\vspace{0.1cm}
{\footnotesize
\begin{tabular}{lcc}  \hline\hline
Decay modes~~~~~~~~~~~~& $A(T_{b3}\to T_{8}\ell^+\ell^-)$  \\\hline
{\color{blue}\bf $T_{b3}\to T_{8}\ell^+\ell^- $ via the $b\to s\ell^+\ell^- $ transition:}\\
$\Lambda^0_b\to \Lambda^0 \ell^+\ell^- $&$-2\lambda_{bs}~E/\sqrt{6}$\\
$\Lambda^0_b\to \Sigma^0 \ell^+\ell^- $& $0$\\
$\Xi^0_b\to \Xi^0 \ell^+\ell^- $& $-\lambda_{bs}~E$\\
$\Xi^-_b\to \Xi^- \ell^+\ell^- $& $\lambda_{bs}~E$\\
{\color{blue}\bf $T_{b3}\to T_{8}\ell^+\ell^- $ via the $b\to d\ell^+\ell^- $ transition:}\\
$\Lambda^0_b\to n \ell^+\ell^- $& $\lambda_{bd}~E$\\
$\Xi^0_b\to \Lambda^0 \ell^+\ell^- $&$-\lambda_{bd}~E/\sqrt{6}$ \\
$\Xi^0_b\to \Sigma^0 \ell^+\ell^- $&$-\lambda_{bd}~E/\sqrt{2}$ \\
$\Xi^-_b\to \Sigma^- \ell^+\ell^- $&$\lambda_{bd}~E$ \\
\hline
\end{tabular}\label{Tab:SU3HATb3}}\vspace{1.5cm}
%
\renewcommand\arraystretch{1.3}
\tabcolsep 0.1in
\centering
\caption{Branching ratios for  $T_{b3}\to T_{8}\ell^+\ell^-$ decays with $1\sigma$ error  in the whole $q^2$ region within $S_1$ and  $S_2$  cases.}\vspace{0.1cm}
{\footnotesize
\begin{tabular}{lcccc}  \hline\hline
Decay modes~~~~~~~~~~~~& Experimental data \cite{PDG2020} &  Our results in $S_1$ &  Our results in $S_2$ & Other predictions\\\hline
$\mathcal{B}(\Lambda^0_b\to \Lambda^0\mu^+\mu^-)(\times10^{-6})$&$1.08\pm 0.28$&$1.08\pm 0.28$&$1.08\pm 0.28$ &$1.05 $ \cite{Faustov:2020thr}\\
$\mathcal{B}(\Xi^0_b\to \Xi^0\mu^+\mu^-)(\times10^{-6})$&$\cdots$&$1.55^{+0.45}_{-0.43}$&$1.77^{+0.49}_{-0.53}$ \\
$\mathcal{B}(\Xi^-_b\to \Xi^-\mu^+\mu^-)(\times10^{-6})$&$\cdots$&$1.65^{+0.49}_{-0.46}$&$1.87^{+0.56}_{-0.54}$ \\ \hline
$\mathcal{B}(\Lambda^0_b\to \Lambda^0\tau^+\tau^-)(\times10^{-7})$&$\cdots$&$2.30\pm0.60$&$2.74^{+0.85}_{-0.71}$&$2.60$ \cite{Faustov:2020thr} \\
$\mathcal{B}(\Xi^0_b\to \Xi^0\tau^+\tau^-)(\times10^{-7})$&$\cdots$&$3.23^{+0.94}_{-0.89}$&$4.42^{+1.36}_{-1.21}$ \\
$\mathcal{B}(\Xi^-_b\to \Xi^-\tau^+\tau^-)(\times10^{-7})$&$\cdots$&$3.42^{+1.01}_{-0.95}$&$4.76^{+1.44}_{-1.36}$ \\\hline
{\color{blue}\bf $b\to d\ell^+\ell^-:$}&\\
$\mathcal{B}(\Lambda^0_b\to n\mu^+\mu^-)(\times10^{-8})$&$\cdots$&$8.15^{+2.44}_{-2.30}$&$7.77^{+2.42}_{-2.28}$ &$^{\big(4.1^{+5.4}_{-1.2}\big)}_{~~3.75} $ \cite{Liu:2019rpm,Faustov:2020thr}\\
$\mathcal{B}(\Xi^0_b\to \Lambda^0\mu^+\mu^-)(\times10^{-8})$&$\cdots$&$1.34^{+0.43}_{-0.39}$ &$1.45^{+0.44}_{-0.45}$\\
$\mathcal{B}(\Xi^0_b\to \Sigma^0\mu^+\mu^-)(\times10^{-8})$&$\cdots$&$3.77^{+1.22}_{-1.10}$ &$4.13^{+1.36}_{-1.24}$\\
$\mathcal{B}(\Xi^-_b\to \Sigma^-\mu^+\mu^-)(\times10^{-8})$&$\cdots$&$8.00^{+2.56}_{-2.40}$&$8.61^{+3.06}_{-2.52}$ \\\hline
$\mathcal{B}(\Lambda^0_b\to n\tau^+\tau^-)(\times10^{-8})$&$\cdots$&$2.07^{+0.62}_{-0.58}$&$2.46^{+0.78}_{-0.70}$&$^{\big(2.9^{+3.7}_{-0.8}\big)}_{~~1.21} $ \cite{Liu:2019rpm,Faustov:2020thr}\\
$\mathcal{B}(\Xi^0_b\to \Lambda^0\tau^+\tau^-)(\times10^{-9})$&$\cdots$&$3.42^{+1.11}_{-1.00}$&$4.63^{+1.71}_{-1.35}$ \\
$\mathcal{B}(\Xi^0_b\to \Sigma^0\tau^+\tau^-)(\times10^{-9})$&$\cdots$&$8.97^{+2.91}_{-2.62}$&$12.23^{+4.12}_{-3.62}$ \\
$\mathcal{B}(\Xi^-_b\to \Sigma^-\tau^+\tau^-)(\times10^{-8})$&$\cdots$&$1.91^{+0.61}_{-0.57}$&$2.60^{+0.87}_{-0.77}$ \\
\hline
\end{tabular}\label{Tab:BrallTb32T8}}
\end{table}

\begin{sidewaystable}
\renewcommand\arraystretch{0.94}
\tabcolsep 0.01in
\centering
\caption{Branching ratios for $T_{b3}\to T_{8}\mu^+\mu^-$ weak decays  in different $q^2$ bins  with $1\sigma$ error  in $S_1$ and $S_2$ cases ( in unit of $10^{-7}$).  }\vspace{0.1cm}
{\footnotesize
\begin{tabular}{c|c|c|c|c|c|c|c|c|c|c|c|c}  \hline\hline
$[q^2_{min},q^2_{max}](\mbox{GeV}^2)$&$[0.1,2.0]$ &$[2.0,4.3]$&$[0.1,4.3]$&$[4.0,6.0]$ &$[1.0,6.0]$&$[6.0,8.0]$&$[4.3,8.68]$ &$[10.09,12.86]$ &$[14.18,16.0]$ &$[0.1,16.0]$ &$[18.0,20.0]$ &$[15.0,20.0]$     \\\hline
$\mathcal{B}(\Lambda^0_b\to \Lambda^0\mu^+\mu^-)^{Exp.}$&$0.71\pm0.27$&$0.28^{+0.28}_{-0.21}$&$2.7\pm2.7$&$0.04^{+0.18}_{-0.02}$&$0.47^{+0.31}_{-0.27}$&$0.50^{+0.26}_{-0.24}$&$0.5\pm0.7$&$2.2\pm0.6$&$1.7\pm0.5$&$7.0\pm2.9$&$2.44\pm0.57$&$6.0\pm1.3$\\
$\mathcal{B}(\Xi^0_b\to \Xi^0\mu^+\mu^-)_{S_1}$&$1.03^{+0.42}_{-0.41}$&$0.41^{+0.43}_{-0.31}$&$3.91^{+4.00}_{-3.85}$&$0.058^{+0.270}_{-0.030}$&$0.68^{+0.47}_{-0.40}$&$0.73^{+0.40}_{-0.36}$&$0.73^{+1.06}_{-1.02}$&$3.21^{+0.97}_{-0.92}$&$2.47^{+0.80}_{-0.77}$&$10.18^{+4.53}_{-4.35}$&$3.23^{+0.85}_{-0.81}$&$8.46^{+2.12}_{-2.01}$\\
$\mathcal{B}(\Xi^0_b\to \Xi^0\mu^+\mu^-)_{S_2}$&$1.14^{+0.49}_{-0.46}$&$0.45^{+0.48}_{-0.34}$&$4.39^{+4.58}_{-4.32}$&$0.065^{+0.302}_{-0.033}$&$0.76^{+0.54}_{-0.45}$&$0.81^{+0.45}_{-0.39}$&$0.81^{+1.20}_{-0.81}$&$3.60^{+1.14}_{-1.06}$&$2.83^{+0.91}_{-0.89}$&$11.46^{+5.29}_{-4.96}$&$3.82^{+1.01}_{-1.03}$&$9.81^{+2.45}_{-2.55}$\\
$\mathcal{B}(\Xi^-_b\to \Xi^-\mu^+\mu^-)_{S_1}$&$1.09^{+0.46}_{-0.43}$&$0.43^{+0.45}_{-0.32}$&$4.15^{+4.28}_{-4.08}$&$0.062^{+0.289}_{-0.032}$&$0.72^{+0.51}_{-0.42}$&$0.77^{+0.43}_{-0.38}$&$0.77^{+1.13}_{-1.09}$&$3.40^{+1.04}_{-0.99}$&$2.62^{+0.86}_{-0.82}$&$10.80^{+4.89}_{-4.65}$&$3.40^{+0.91}_{-0.87}$&$8.94^{+2.30}_{-2.15}$\\
$\mathcal{B}(\Xi^-_b\to \Xi^-\mu^+\mu^-)_{S_2}$&$1.23^{+0.52}_{-0.49}$&$0.48^{+0.51}_{-0.37}$&$4.57^{+4.92}_{-4.50}$&$0.069^{+0.324}_{-0.035}$&$0.81^{+0.58}_{-0.47}$&$0.86^{+0.50}_{-0.42}$&$0.86^{+1.28}_{-0.86}$&$3.84^{+1.17}_{-1.13}$&$3.00^{+0.99}_{-0.94}$&$12.22^{+5.56}_{-5.38}$&$3.98^{+1.14}_{-1.03}$&$10.30^{+2.77}_{-2.56}$\\  \hline
$\mathcal{B}(\Lambda^0_b\to n\mu^+\mu^-)_{S_1}$&$0.047^{+0.020}_{-0.019}$&$0.019^{+0.020}_{-0.014}$&$0.180^{+0.189}_{-0.177}$&$0.0027^{+0.0127}_{-0.0014}$&$0.031^{+0.022}_{-0.019}$&$0.034^{+0.019}_{-0.017}$&$0.034^{+0.050}_{-0.047}$&$0.152^{+0.048}_{-0.045}$&$0.123^{+0.041}_{-0.039}$&$0.482^{+0.225}_{-0.208}$&$0.236^{+0.065}_{-0.062}$&$0.491^{+0.128}_{-0.121}$\\
$\mathcal{B}(\Lambda^0_b\to n\mu^+\mu^-)_{S_2}$&$0.044^{+0.020}_{-0.018}$&$0.017^{+0.019}_{-0.013}$&$0.169^{+0.176}_{-0.166}$&$0.0025^{+0.0116}_{-0.0012}$&$0.029^{+0.021}_{-0.017}$&$0.030^{+0.017}_{-0.015}$&$0.031^{+0.045}_{-0.031}$&$0.132^{+0.042}_{-0.040}$&$0.107^{+0.036}_{-0.034}$&$0.423^{+0.204}_{-0.181}$&$0.215^{+0.059}_{-0.056}$&$0.445^{+0.117}_{-0.114}$\\
$\mathcal{B}(\Xi^0_b\to \Lambda^0\mu^+\mu^-)_{S_1}$&$0.008\pm0.003$&$0.003\pm0.003$&$0.029^{+0.031}_{-0.029}$&$0.0000^{+0.0025}_{-0.0000}$&$0.005^{+0.004}_{-0.003}$&$0.005\pm0.003$&$0.005\pm0.008$&$0.025\pm0.008$&$0.020\pm0.007$&$0.079^{+0.038}_{-0.035}$&$0.039^{+0.012}_{-0.011}$&$0.081^{+0.023}_{-0.021}$\\
$\mathcal{B}(\Xi^0_b\to \Lambda^0\mu^+\mu^-)_{S_2}$&$0.008^{+0.004}_{-0.003}$&$0.003^{+0.003}_{-0.002}$&$0.030^{+0.034}_{-0.029}$&$0.0004^{+0.0021}_{-0.0002}$&$0.005^{+0.004}_{-0.003}$&$0.005\pm0.003$&$0.005^{+0.008}_{-0.005}$&$0.024\pm0.008$&$0.020^{+0.007}_{-0.006}$&$0.077^{+0.039}_{-0.034}$&$0.041^{+0.012}_{-0.011}$&$0.083^{+0.024}_{-0.022}$\\
$\mathcal{B}(\Xi^0_b\to \Sigma^0\mu^+\mu^-)_{S_1}$&$0.023^{+0.010}_{-0.009}$&$0.009^{+0.010}_{-0.007}$&$0.087^{+0.093}_{-0.085}$&$0.0013^{+0.0062}_{-0.0013}$&$0.015^{+0.011}_{-0.009}$&$0.016^{+0.010}_{-0.008}$&$0.016^{+0.025}_{-0.023}$&$0.073^{+0.024}_{-0.023}$&$0.058^{+0.022}_{-0.019}$&$0.231^{+0.113}_{-0.103}$&$0.103^{+0.031}_{-0.028}$&$0.225^{+0.063}_{-0.057}$\\
$\mathcal{B}(\Xi^0_b\to \Sigma^0\mu^+\mu^-)_{S_2}$&$0.024^{+0.011}_{-0.010}$&$0.010^{+0.011}_{-0.007}$&$0.093^{+0.101}_{-0.092}$&$0.0014^{+0.0065}_{-0.0007}$&$0.016^{+0.012}_{-0.009}$&$0.017^{+0.010}_{-0.009}$&$0.017^{+0.025}_{-0.017}$&$0.075^{+0.025}_{-0.023}$&$0.060^{+0.023}_{-0.020}$&$0.237^{+0.120}_{-0.103}$&$0.114^{+0.034}_{-0.032}$&$0.245^{+0.069}_{-0.064}$\\
$\mathcal{B}(\Xi^-_b\to \Sigma^-\mu^+\mu^-)_{S_1}$&$0.048^{+0.022}_{-0.020}$&$0.019^{+0.021}_{-0.014}$&$0.184^{+0.200}_{-0.181}$&$0.0027^{+0.0133}_{-0.0014}$&$0.032^{+0.024}_{-0.019}$&$0.034^{+0.021}_{-0.017}$&$0.034^{+0.053}_{-0.049}$&$0.155^{+0.053}_{-0.047}$&$0.124^{+0.045}_{-0.041}$&$0.491^{+0.240}_{-0.218}$&$0.220^{+0.066}_{-0.061}$&$0.478^{+0.140}_{-0.122}$\\
$\mathcal{B}(\Xi^-_b\to \Sigma^-\mu^+\mu^-)_{S_2}$&$0.051^{+0.024}_{-0.020}$&$0.020^{+0.023}_{-0.015}$&$0.196^{+0.213}_{-0.193}$&$0.0029^{+0.0140}_{-0.0015}$&$0.034^{+0.026}_{-0.020}$&$0.036^{+0.021}_{-0.018}$&$0.036^{+0.054}_{-0.036}$&$0.159^{+0.056}_{-0.049}$&$0.129^{+0.047}_{-0.043}$&$0.508^{+0.260}_{-0.225}$&$0.243^{+0.077}_{-0.066}$&$0.515^{+0.151}_{-0.130}$\\\hline
\end{tabular}\label{Tab:BrbinTb32T8mu}}
\renewcommand\arraystretch{0.94}
\tabcolsep 0.15in
\centering
\caption{Branching ratios for $T_{b3}\to T_{8}\tau^+\tau^-$ weak decays  in different $q^2$ bins with $1\sigma$ error  in $S_1$ and $S_2$ cases ( in unit of $10^{-7}$). }\vspace{0.1cm}
{\footnotesize
\begin{tabular}{ccccccccc}  \hline\hline
$[q^2_{min},q^2_{max}](\mbox{GeV}^2)$&$[14.18,16.0]$ in $S_1$ &$[14.18,16.0]$ in $S_2$&$[18.0,20.0]$ in $S_1$&$[18.0,20.0]$ in $S_2$  &$[15.0,20.0]$ in $S_1$   &$[15.0,20.0]$ in $S_2$   \\\hline
$\mathcal{B}(\Lambda^0_b\to \Lambda^0\tau^+\tau^-)$&$0.83\pm0.25$&$0.84^{+0.27}_{-0.25}$&$1.51\pm0.36$&$1.52\pm0.37$&$3.41\pm0.76$&$3.44^{+0.86}_{-0.80}$ \\
$\mathcal{B}(\Xi^0_b\to \Xi^0\tau^+\tau^-)$&$1.21^{+0.39}_{-0.38}$&$1.40^{+0.46}_{-0.45}$&$2.00^{+0.53}_{-0.50}$&$2.32^{+0.66}_{-0.59}$&$4.79^{+1.20}_{-1.14}$&$5.65^{+1.43}_{-1.44}$ \\
$\mathcal{B}(\Xi^-_b\to \Xi^-\tau^+\tau^-)$&$1.29^{+0.42}_{-0.40}$&$1.49^{+0.50}_{-0.48}$&$2.11^{+0.57}_{-0.54}$&$2.50\pm0.67$&$5.07^{+1.30}_{-1.22}$&$5.92^{+1.58}_{-1.48}$ \\\hline
$\mathcal{B}(\Lambda^0_b\to n\tau^+\tau^-)$&$0.060^{+0.020}_{-0.019}$&$0.053\pm0.018$&$0.147^{+0.041}_{-0.039}$&$0.133^{+0.038}_{-0.034}$&$0.282^{+0.074}_{-0.070}$&$0.257^{+0.068}_{-0.064}$\\
$\mathcal{B}(\Xi^0_b\to \Lambda^0\tau^+\tau^-)$&$0.010^{+0.004}_{-0.003}$&$0.0095^{+0.0039}_{-0.0030}$&$0.024\pm0.007$&$0.026\pm0.007$&$0.047^{+0.013}_{-0.012}$&$0.048^{+0.014}_{-0.013}$ \\
$\mathcal{B}(\Xi^0_b\to \Sigma^0\tau^+\tau^-)$&$0.029^{+0.011}_{-0.009}$&$0.030^{+0.011}_{-0.010}$&$0.064^{+0.019}_{-0.017}$&$0.071^{+0.021}_{-0.019}$&$0.129^{+0.036}_{-0.033}$&$0.140^{+0.041}_{-0.037}$ \\
$\mathcal{B}(\Xi^-_b\to \Sigma^-\tau^+\tau^-)$&$0.061^{+0.022}_{-0.020}$&$0.063^{+0.023}_{-0.021}$&$0.137^{+0.041}_{-0.038}$&$0.152^{+0.047}_{-0.041}$&$0.273^{+0.080}_{-0.070}$&$0.301^{+0.086}_{-0.078}$ \\\hline
\end{tabular}\label{Tab:BrbinTb32T8tau}}
\renewcommand\arraystretch{0.94}
\tabcolsep 0.01in
\centering
\caption{Longitudinal polarization fractions and forward-backward asymmetries for $\Lambda^0_b\to \Lambda^0\mu^+\mu^-,\Lambda^0\tau^+\tau^-$  decays  in different $q^2$ bins  with $1\sigma$ error  in  $S_2$ case. }\vspace{0.1cm}
{\footnotesize
\begin{tabular}{c|c|c|c|c|c|c|c|c|c|c|c|c|c}  \hline\hline
$[q^2_{min},q^2_{max}](\mbox{GeV}^2)$&$[0.1,2.0]$ &$[2.0,4.3]$&$[0.1,4.3]$&$[4.0,6.0]$ &$[1.0,6.0]$&$[6.0,8.0]$&$[4.3,8.68]$ &$[10.09,12.86]$ &$[14.18,16.0]$ &$[0.1,16.0]$ &$[18.0,20.0]$ &$[15.0,20.0]$ &whole $q^2$ regions    \\\hline
$\overline{f_L}(\Lambda^0_b\to \Lambda^0\mu^+\mu^-)$&$0.64^{+0.04}_{-0.01}$&$0.86\pm0.01$&$0.77^{+0.06}_{-0.03}$&$0.81\pm0.01$&$0.83^{+0.02}_{-0.01}$&$0.73\pm0.01$&$0.77\pm0.02$&$0.57^{+0.00}_{-0.02}$&$0.46\pm0.01$&$0.66^{+0.02}_{-0.03}$&$0.36\pm0.01$&$0.39^{+0.02}_{-0.01}$&$0.60\pm0.02$\\
$\langle f_L\rangle(\Lambda^0_b\to \Lambda^0\mu^+\mu^-)$&$0.36^{+0.04}_{-0.02}$&$0.86^{+0.06}_{-0.04}$&$0.42^{+0.04}_{-0.02}$&$0.81^{+0.03}_{-0.02}$&$0.82^{+0.05}_{-0.02}$&$0.73\pm0.01$&$0.75\pm0.03$&$0.56\pm0.01$&$0.45\pm0.01$&$0.47^{+0.04}_{-0.02}$&$0.36\pm0.01$&$0.40^{+0.01}_{-0.02}$&$0.34^{+0.03}_{-0.02}$\\
$\overline{A^{\ell}_{FB}}(\Lambda^0_b\to \Lambda^0\mu^+\mu^-)$&$0.12\pm0.01$&$0.05^{+0.00}_{-0.01}$&$0.08^{+0.00}_{-0.01}$&$-0.05^{+0.00}_{-0.01}$&$0.03^{+0.00}_{-0.01}$&$-0.15^{+0.00}_{-0.01}$&$-0.12^{+0.00}_{-0.01}$&$-0.29\pm0.01$&$-0.36\pm0.00$&$-0.15\pm0.01$&$-0.31^{+0.01}_{-0.00}$&$-0.35^{+0.01}_{-0.02}$&$-0.19^{+0.00}_{-0.01}$\\
$\langle A^{\ell}_{FB}\rangle(\Lambda^0_b\to \Lambda^0\mu^+\mu^-)$&$0.08^{+0.00}_{-0.01}$&$0.06^{+0.00}_{-0.01}$&$0.07\pm0.00$&$-0.05^{+0.00}_{-0.01}$&$0.06^{+0.00}_{-0.01}$&$-0.15^{+0.00}_{-0.01}$&$-0.12^{+0.00}_{-0.01}$&$-0.29^{+0.01}_{-0.02}$&$-0.37^{+0.02}_{-0.01}$&$-0.03^{+0.00}_{-0.01}$&$-0.30\pm0.01$&$-0.34^{+0.01}_{-0.02}$&$-0.04^{+0.00}_{-0.01}$\\\hline
$\overline{f_L}(\Lambda^0_b\to \Lambda^0\tau^+\tau^-)$&&&&&&&&&$0.46^{+0.00}_{-0.01}$&&$0.36^{+0.01}_{-0.00}$&$0.39^{+0.02}_{-0.01}$&$0.42\pm0.01$ \\
$\langle f_L\rangle(\Lambda^0_b\to \Lambda^0\tau^+\tau^-)$&&&&&&&&&$0.46\pm0.01$&&$0.36\pm0.01$&$0.39^{+0.02}_{-0.01}$&$0.40^{+0.03}_{-0.01}$\\
$\overline{A^{\ell}_{FB}}(\Lambda^0_b\to \Lambda^0\tau^+\tau^-)$&&&&&&&&&$-0.36\pm0.00$&&$-0.31\pm0.01$&$-0.34^{+0.01}_{-0.02}$&$-0.34\pm0.01$ \\
$\langle A^{\ell}_{FB}\rangle(\Lambda^0_b\to \Lambda^0\tau^+\tau^-)$&&&&&&&&&$-0.36^{+0.01}_{-0.02}$&&$-0.30\pm0.01$&$-0.34^{+0.01}_{-0.02}$&$-0.33^{+0.02}_{-0.01}$ \\
\hline
\end{tabular}\label{Tab:fLAlFBTb32T81}}
\end{sidewaystable}

Among $\Lambda^0_b\to \Lambda^0 \ell^+\ell^-$, $\Xi^0_b\to \Xi^0 \ell^+\ell^- $ $\Xi^-_b\to \Xi^- \ell^+\ell^-$, $\Lambda^0_b\to n \ell^+\ell^-$, $\Xi^0_b\to \Lambda^0 \ell^+\ell^-$, $\Xi^0_b\to \Sigma^0 \ell^+\ell^-$, and $\Xi^-_b\to \Sigma^- \ell^+\ell^-$ decays, only $\Lambda^0_b\to \Lambda^0 \mu^+\mu^-$ decay has been measured, and its branching ratios in the whole $q^2$ region and in different $q^2$ bins are listed in Tab. \ref{Tab:BrallTb32T8} and Tab. \ref{Tab:BrbinTb32T8mu}, respectively. One can constrain  the relevant SU(3) flavor parameters by the experimental data within $1\sigma$ error bar and then predict other not-yet-measured branching ratios.
 Two cases will be considered in our analysis of $T_{b3}\to T_8\ell^+\ell^-$ decays.

\begin{itemize}
\item[\bf$S_1$ :] {\bf The SU(3) flavor symmetry  parameters
without the baryonic momentum-transfer $q^2$  dependence.} We treat the SU(3) flavor parameters $(E)^{L(R),s_p,s_k}_{VA,\lambda}(q^2)$ as constants without $q^2$ dependence, which will  lead  $H_M(q^2)$ in Eq. (\ref{EQ:dB}) to a constant, too.
 We  use the $1\sigma$ error experimental data of $\mathcal{B}(\Lambda^0_b\to \Lambda^0 \mu^+\mu^-)$ to constrain $H_M(q^2)$ ($i.e.$, $|E|^2$), and then predict  other $\mathcal{B}(T_{b3}\to T_8\ell^+\ell^-)$ by the amplitude relations in Tab. \ref{Tab:SU3HATb3}.

\item[\bf $S_2$:]{\bf  The SU(3) flavor symmetry  parameters  with the baryonic momentum-transfer $q^2$  dependence.}  In order to obtain more precise predictions, we use
the hadronic helicity amplitude expressions  in Eq. (\ref{EQ:H}), which are $q^2$ dependent and can be expressed by  the Wilson coefficients and the form factors.
The expressions of the Wilson coefficients  without the LD contributions are taken from Ref. \cite{Buras:1994dj}.
As for the $q^2$ dependent form factors involving the
$T_{b3}\to T_8$ transitions,    we  use the recent lattice QCD results of $\Lambda^0_b\to \Lambda^0$ \cite{Detmold:2016pkz}, in which  the  form factors are  parametrized by
 \begin{eqnarray}
f(q^2)=\frac{f(0)}{1-q^2/(m^f_{pole})^2}\left[1+\frac{a^f_1}{a^f_0}z(q^2)+\frac{a^f_2}{a^f_0}[z(q^2)]^2\right],
\end{eqnarray}
where $f=f_+,f_\perp,f_0,g_+,g_\perp,g_0,h_+,h_\perp,\widetilde{h}_+,\widetilde{h}_\perp$, and the details of $z(q^2)$ and $m^f_{pole}$ can be found in Ref. \cite{Detmold:2016pkz}.  We keep $f_+(0)$ as an undetermined  constant without $q^2$ dependence, and other $f(0)$ can be expressed as  $\frac{a^{f}_0 }{a^{f_+}_0}f_+(0)$. The central values of $a^f_i$ in Tab. V of Ref. \cite{Detmold:2016pkz} will be used in our analysis.
Since these form factors also preserve the SU(3) flavor symmetry, the same relations in Tab. \ref{Tab:SU3HATb3} will be used for $f_+(0)$.  We  use the $1\sigma$ error experimental data of $\mathcal{B}(\Lambda^0_b\to \Lambda^0 \mu^+\mu^-)$ to constrain $f_+(0)$ and then predict other $\mathcal{B}(T_{b3}\to T_8\ell^+\ell^-)$.

\end{itemize}

Using the experimental data of $\mathcal{B}(\Lambda^0_b\to \Lambda^0\mu^+\mu^-)$ in the whole $q^2$ region, one can obtain the branching ratios for $T_{b3}\to T_{8}\mu^+\mu^-$ and $T_{b3}\to T_{8}\tau^+\tau^-$ weak decays  in the whole $q^2$ region, which are listed in the third and forth columns of Tab. \ref{Tab:BrallTb32T8} for $S_1$ case and $S_2$ case, respectively.
Noted that, the amplitude relations listed in Tab. \ref{Tab:SU3HATb3} are obtained from
the SU(3) flavor symmetry; nevertheless, the different baryon masses in the same baryon multiplets are considered in
the branching ratio predictions, and the below is same.

Previous  predictions are also listed in the last column of Tab. \ref{Tab:BrallTb32T8} for comparing.
Since the results of $T_{b3}\to T_{8}e^+e^-$ decays  are quite similar to ones of $T_{b3}\to T_{8}\mu^+\mu^-$ decays, we   only show $T_{b3}\to T_{8}\mu^+\mu^-$ in this work.
We have the following remarks for the results in Tab. \ref{Tab:BrallTb32T8}.
\begin{itemize}
\item  Comparing the branching ratios in $S_1$ and $S_2$ cases, one can see that the predictions are slightly different  between $S_1$  and $S_2$ cases, which are mainly due to the $q^2$ dependence of the  hadronic helicity amplitudes.

\item Comparing our predictions for $\mathcal{B}(\Lambda^0_b\to \Lambda^0\tau^+\tau^-),~\mathcal{B}(\Lambda^0_b\to n\tau^+\tau^-)$ and $\mathcal{B}(\Lambda^0_b\to n\mu^+\mu^-)$ with previous ones in the
relativistic quark model \cite{Faustov:2020thr} and the Bethe-Salpeter equation approach \cite{Liu:2019rpm}, one can see that our predicted
$\mathcal{B}(\Lambda^0_b\to \Lambda^0\tau^+\tau^-)$ and $\mathcal{B}(\Lambda^0_b\to n\mu^+\mu^-)$  are quite consistent with previous ones,  nevertheless, our central value of $\mathcal{B}(\Lambda^0_b\to n\tau^+\tau^-)$ is about 2 times larger than theirs.

\item  Many  branching ratio predictions for $T_{b3}\to T_{8}\ell^+\ell^-$ are obtained for the first time.  The not-yet-measured  $\mathcal{B}(T_{b3}\to T_{8}\ell^+\ell^-)$ are on the order of  $\mathcal{O}(10^{-8}-10^{-6})$, and some of them  could be reached by the LHCb or Belle-II experiments.

\end{itemize}

Using the experimental data of $\mathcal{B}(\Lambda^0_b\to \Lambda^0\mu^+\mu^-)$ in different $q^2$ bins, one can get  the branching ratios of $T_{b3}\to T_{8}\mu^+\mu^-$ and $T_{b3}\to T_{8}\tau^+\tau^-$ weak decays  in different $q^2$ bins within $S_1$   and  $S_2$ cases, which are collected for reference in Tab. \ref{Tab:BrbinTb32T8mu} and  Tab. \ref{Tab:BrbinTb32T8tau}, respectively.

The longitudinal polarization fractions and the
leptonic forward-backward asymmetries  with two ways of integration for  $T_{b3}\to T_{8}\ell^+\ell^-$ decays could also be obtained in the $S_2$ case. As shown in Eq. (\ref{Eq:fLc}) and Eq. (\ref{Eq:AlFBc}),  the $N(q^2)$ terms  are canceled in the ratios; therefore, the longitudinal polarization fractions and the
leptonic forward-backward asymmetries  only depend on the hadronic helicity amplitudes, which preserve the SU(3) flavor symmetry in $T_{b3}\to T_{8}\ell^+\ell^-$ decays.  So  the longitudinal polarization fractions and the
leptonic forward-backward asymmetries of  all $T_{b3}\to T_{8}\mu^+\mu^-$ ($T_{b3}\to T_{8}\tau^+\tau^-$) decays are very similar to each other in certain $q^2$ bins. We take ones of $\Lambda^0_b\to \Lambda^0\mu^+\mu^-,\Lambda^0\tau^+\tau^-$ as  examples, which are given in Tab. \ref{Tab:fLAlFBTb32T81}. Excepting in $[0.1,2.0]$, $[0.1,4.3]$, $[0.1,16.0]$ and the whole $q^2$ regions, $\overline{f_L}$ and $\langle f_L\rangle$  ($\overline{A^{\ell}_{FB}}$ and $\langle A^{\ell}_{FB}\rangle$) with different $q^2$ integration ways are quite similar in  other certain $q^2$ bins.
So the obvious differentiation between $\overline{f_L}$ and $\langle f_L\rangle$  ($\overline{A^{\ell}_{FB}}$ and $\langle A^{\ell}_{FB}\rangle$) mainly appears in the quite low $q^2$ regions.
Note that the normalized longitudinal polarization fraction and
normalized leptonic forward-backward asymmetry of $\Lambda^0_b\to \Lambda^0\mu^+\mu^-$ in $q^2\in[15,20]$ GeV$^2$  have been  measured by the LHCb experiment \cite{PDG2020}, 
\begin{eqnarray}
\langle f_L\rangle(\Lambda^0_b\to \Lambda^0\mu^+\mu^-)_{[15,20]}&=&0.61^{+0.11}_{-0.14},\nonumber\\
\langle A^\ell_{FB}\rangle(\Lambda^0_b\to \Lambda^0\mu^+\mu^-)_{[15,20]}&=&-0.39\pm0.04\pm0.01.\label{Eq:ExpfLAlFB15t18}
\end{eqnarray}
We do not impose the above experimental bounds  but leave
them as predictions.
Comparing with the experimental results for  $\langle f_L\rangle(\Lambda^0_b\to \Lambda^0\mu^+\mu^-)_{[15,20]}$ and $\langle A^\ell_{FB}\rangle(\Lambda^0_b\to \Lambda^0\mu^+\mu^-)_{[15,20]}$,   our prediction of  $\langle f_L\rangle(\Lambda^0_b\to \Lambda^0\mu^+\mu^-)_{[15,20]}$  and $\langle A^\ell_{FB}\rangle(\Lambda^0_b\to \Lambda^0\mu^+\mu^-)_{[15,20]}$   are agreeable with their experimental data within $1.5\sigma$   and  $1\sigma$ error ranges, respectively.

In addition, the lepton flavor universality $R_{T_{b3}\to T_8}$ in three $q^2$ bins within the $S_2$ case are given in Tab. \ref{Tab:RbinTb32T8mue}.  One can see that all predictions  in three  $q^2$  bins are virtually
indistinguishable from unity; $i.e.$, the lepton mass effects on  all $R_{\mathcal{B}_1\to \mathcal{B}_2}$ are small in both the low-$q^2$ region and high-$q^2$ region in the SM.
\begin{table}[ht]
\renewcommand\arraystretch{1.3}
\tabcolsep 0.35in
\centering
\caption{Lepton flavor universality of  $T_{b3}\to T_8\ell^+\ell^-$ baryon weak decays   in different $q^2$ bins with $1\sigma$ error  in the $S_2$ case. }\vspace{0.1cm}
{\footnotesize
\begin{tabular}{llll}  \hline\hline
$[q^2_{min},q^2_{max}](\mbox{GeV}^2)$&$[1,6]$  &$[0.1,16.0]$       &$[15.0,20.0]$   \\\hline
$R_{\Lambda^0_b\to \Lambda^0}$&$0.99^{+0.05}_{-0.03}$    &$0.98^{+0.07}_{-0.04}$  &$1.02^{+0.04}_{-0.07}$ \\
$R_{\Xi^0_b\to \Xi^0}$        &$0.99\pm0.04$             &$0.99^{+0.06}_{-0.05}$  &$0.99^{+0.08}_{-0.06}$ \\
$R_{\Xi^-_b\to \Xi^-}$        &$0.99\pm0.04$             &$0.99^{+0.06}_{-0.05}$  &$1.03^{+0.03}_{-0.09}$ \\\hline
$R_{\Lambda^0_b\to n}$        &$1.00\pm0.03$             &$0.98^{+0.07}_{-0.04}$  &$1.01^{+0.03}_{-0.05}$\\
$R_{\Xi^0_b\to \Lambda^0}$    &$1.01^{+0.02}_{-0.05}$    &$1.01^{+0.04}_{-0.08}$  &$1.00\pm0.04$ \\
$R_{\Xi^0_b\to \Sigma^0}$     &$0.98^{+0.04}_{-0.03}$    &$0.98^{+0.07}_{-0.04}$  &$1.00\pm0.03$ \\
$R_{\Xi^-_b\to \Sigma^-}$     &$1.00^{+0.03}_{-0.04}$    &$1.03^{+0.02}_{-0.09}$  &$1.00^{+0.03}_{-0.02}$ \\\hline
\end{tabular}\label{Tab:RbinTb32T8mue}}
\end{table}

\subsection{$T_{c3}$ semileptonic weak decays}
Similar to $T_{c3,8}\to T'_{8}\gamma$ radiative decays \cite{Singer:1995is,Verma:1988gf,Lach:1995we,Azimov:1996uf}, $T_{c3}\to T_{8}\ell^+\ell^-$ decays receive  single-quark,   two-quark, and three-quark
transition contributions with the $W$-exchange  and   internal radiation contributions.
 The internal radiation contributions are suppressed by the two $W$ propagators and  can be safely neglected.
The  SU(3) IRA  hadronic helicity amplitudes for $T_{c3}\to T_{8}\ell^+\ell^-$  may be parametrized as
\begin{eqnarray}
H(T_{c3} \to T_8\ell^+\ell^-)^{L(R),s_p,s_k}_{VA,\lambda}&=&f_1(T_{c3})^{[ij]}T'(\bar{3})^k(T_8)_{[ij]k}+f_2(T_{c3})^{[ij]}T'(3)^k(T_8)_{[ik]j}\nonumber\\
&&+\Big(\widetilde{f}_1H(\bar{6})^{lk}_j+\widetilde{f}_4H(15)^{lk}_j\Big)(T_{c3})^{[ij]}(T_8)_{[il]k}\nonumber\\
&&+\Big(\widetilde{f}_2H(\bar{6})^{lk}_j+\widetilde{f}_5H(15)^{lk}_j\Big)(T_{c3})^{[ij]}(T_8)_{[ik]l}\nonumber\\
&&+\Big(\widetilde{f}_3H(\bar{6})^{lk}_j+\widetilde{f}_6H(15)^{lk}_j\Big)(T_{c3})^{[ij]}(T_8)_{[lk]i},\label{Eq:HTc32T8}
\end{eqnarray}
where the SU(3) flavor symmetry  parameters  $f_i\equiv(f_i)^{L(R),s_p,s_k}_{VA,\lambda}$ and   $\widetilde{f}_i\equiv(\widetilde{f}_i)^{L(R),s_p,s_k}_{VA,\lambda}$.  The $f_i$ terms in Eq. (\ref{Eq:HTc32T8}) and the later $g_i$ terms  in Eq. (\ref{Eq:HT82T8p}) denote the short distance (SD)  and the LD contributions via the single-quark transitions.
The $\widetilde{f}_i$ terms in Eq. (\ref{Eq:HTc32T8}) and the later $\widetilde{g}_i$ terms  in Eq. (\ref{Eq:HT82T8p}) denote the $W$-exchange contributions of  the two-quark and three-quark
transitions.  $T'(\bar{3})=(1,0,0)$   denotes the transition operators $(\bar{q}_2c)$  with $q_2=u$, and   $H(\bar{6})^{lk}_j$ ($H(15)^{lk}_j$) related to the $(\bar{q}_lq^j)(\bar{q}_kc)$ operator  is antisymmetric (symmetric) in upper indices.
The nonvanishing $H(\bar{6})^{lk}_j$  and $H(15)^{lk}_j$ for $c\to su\bar{d},du\bar{s},u\bar{d}d,u\bar{s}s$ transitions can be found in Ref. \cite{Wang:2017azm}.
Using $l,k$ antisymmetric in $H(\bar{6})^{lk}_j$ and $l,k$ symmetric in $H(15)^{lk}_j$,  we have
\begin{eqnarray}
\widetilde{f}_2=-\widetilde{f}_1, ~~~\widetilde{f}_5=\widetilde{f}_4, ~~~\widetilde{f}_6=0,
\end{eqnarray}
which will be used in the following discussion.

For the $W$-exchange transitions,
there are three kinds of charm quark decaying  into light quarks
\begin{eqnarray}
d+c\to u+s+\ell^+\ell^-,~~~~~~  d+c\to u+d+\ell^+\ell^-~~(s+c\to u+s+\ell^+\ell^-),~~~~~~~s+c\to u+d+\ell^+\ell^-, \label{Eq:3kcdecays}
\end{eqnarray}
 which are related to
 $H(\bar{6},15)^{13}_2$, $H(\bar{6},15)^{12}_2~\big(H(\bar{6},15)^{13}_3\big)$ and  $H(\bar{6},15)^{12}_3$, and they  are proportional to $V^*_{cs}V_{ud}\approx1$,  $V^*_{cd}V_{ud}~(V^*_{cs}V_{us})\approx-s_c~(s_c)$  and  $V^*_{cd}V_{us}\approx-s_c^2$ with $s_c\equiv sin\theta_C\approx0.22453$,  respectively.  So three kinds   decays given  in Eq. (\ref{Eq:3kcdecays}) are  called Cabibbo
allowed, singly Cabibbo suppressed, and doubly Cabibbo suppressed decays,  respectively.

The SU(3) IRA amplitudes of the  $T_{c3}\to T_{8}\ell^+\ell^- $ weak decays are given in the third column of Tab. \ref{Tab:SU3HATc3}, and for a better understanding, the information of
relevant CKM matrix elements  is also listed in this Table.
\begin{table}[b]
\renewcommand\arraystretch{1.3}
\tabcolsep 0.15in
\centering
\caption{The SU(3) IRA amplitudes of the  $T_{c3}\to T_{8}\ell^+\ell^- $ weak decays in the $S_1$ case,  $F_1\equiv f_1+f_2$, $\widetilde{F}_1\equiv\widetilde{f}_1-\widetilde{f}_3+\widetilde{f}_4$, $\widetilde{F}_2\equiv\widetilde{f}_1-\widetilde{f}_3-\widetilde{f}_4$, and $\widetilde{F}\equiv\widetilde{f}_1-\widetilde{f}_3$.   }\vspace{0.1cm}
{\footnotesize
\begin{tabular}{lcc}  \hline
Decay modes~~~~~~~~~~~~& $A(T_{c3}\to T_{8}\ell^+\ell^-)$ & Approximative $A(T_{c3}\to T_{8}\ell^+\ell^-)$  \\\hline
{\color{blue}\bf  Cabibbo
allowed $T_{c3}\to T_{8}\ell^+\ell^-  $:}\\
$\Lambda^+_c \to \Sigma^+\ell^+\ell^- $ & $-\widetilde{F}_1$& $-\widetilde{F}$\\
$\Xi^0_c \to \Xi^0\ell^+\ell^- $ & $-\widetilde{F}_2$& $-\widetilde{F}$\\
{\color{blue}\bf  singly Cabibbo suppressed $T_{c3}\to T_{8}\ell^+\ell^-  $:}\\
$\Lambda^+_c\to p \ell^+\ell^-  $& $\big[F_1-\big(\frac{5}{8}\widetilde{F}_1-\frac{1}{8}\widetilde{F}_2\big)\big]s_c$& $\big[F_1-\frac{1}{2}\widetilde{F}~\big]s_c$\\
$\Xi^+_c\to \Sigma^+ \ell^+\ell^-  $&$\big[-F_1-\big(\frac{5}{8}\widetilde{F}_1-\frac{1}{8}\widetilde{F}_2\big)\big]s_c$& $-\big[F_1+\frac{1}{2}\widetilde{F}~\big]s_c$\\
$\Xi^0_c\to \Lambda^0 \ell^+\ell^-  $&$\big[F_1+3\big(\frac{1}{8}\widetilde{F}_1-\frac{5}{8}\widetilde{F}_2\big)\big]s_c/\sqrt{6}$ & $\big[F_1-\frac{3}{2}\widetilde{F}~\big]s_c/\sqrt{6}$ \\
$\Xi^0_c\to \Sigma^0 \ell^+\ell^-  $&$\big[-F_1+\big(\frac{1}{8}\widetilde{F}_1-\frac{5}{8}\widetilde{F}_2\big)\big]s_c/\sqrt{2}$& $-\big[F_1+\frac{1}{2}\widetilde{F}~\big]s_c/\sqrt{2}$ \\
{\color{blue}\bf doubly Cabibbo suppressed $T_{c3}\to T_{8}\ell^+\ell^-  $:}\\
$\Xi^+_c\to p\ell^+\ell^- $&$\widetilde{F}_1s^2_c$&$\widetilde{F}s^2_c$\\
$\Xi^0_c\to n\ell^+\ell^- $&$\widetilde{F}_2s^2_c$&$\widetilde{F}s^2_c$\\
\hline
\end{tabular}\label{Tab:SU3HATc3}}\vspace{1cm}
\renewcommand\arraystretch{1.3}
\tabcolsep 0.1in
\centering
\caption{Branching ratios of $T_{c3}\to T_{8}\ell^+\ell^-$ decays  within $1\sigma$ theoretical error in the $S_1$ case.}\vspace{0.1cm}
{\footnotesize
\begin{tabular}{rcccc}  \hline
Decay modes~~~~~~~~~~~~& Exp. UL \cite{PDG2020} &  Our  predictions without $F_1$ & Others without LD & Others with LD \\\hline
$\mathcal{B}(\Lambda^+_c \to \Sigma^+e^+e^- )(\times10^{-6})$ & $\cdots$&$\leq2.63$\\
$\mathcal{B}(\Xi^0_c \to \Xi^0e^+e^- )(\times10^{-6})$ & $\cdots$&$\leq2.35$\\
$\mathcal{B}(\Lambda^+_c\to pe^+e^-)(\times10^{-8})$&$\leq550$&$\leq7.95$&$^{(3.8\pm0.5)\times10^{-4}}_{(4.05\pm2.37)\times10^{-6}}$ \cite{Faustov:2018dkn,Sirvanli:2016wnr}&$^{37\pm8}_{420\pm73}$ \cite{Faustov:2018dkn,Sirvanli:2016wnr}\\
$\mathcal{B}(\Xi^+_c\to \Sigma^+e^+e^-)(\times10^{-7})$&$\cdots$&$\leq1.29$ \\
$\mathcal{B}(\Xi^0_c\to\Lambda^0e^+e^-)(\times10^{-8})$&$\cdots$&$\leq8.69$ \\
$\mathcal{B}(\Xi^0_c\to\Sigma^0e^+e^-)(\times10^{-8})$&$\cdots$&$\leq2.22$ \\
$\mathcal{B}(\Xi^+_c\to pe^+e^-)(\times10^{-8}) $&$\cdots$&$\leq5.55$\\
$\mathcal{B}(\Xi^0_c\to ne^+e^-)(\times10^{-8}) $&$\cdots$&$\leq1.92$\\
\hline
$\mathcal{B}(\Lambda^+_c \to \Sigma^+\mu^+\mu^- )(\times10^{-6})$ &$\cdots$ &$\leq2.50$\\
$\mathcal{B}(\Xi^0_c \to \Xi^0\mu^+\mu^- )(\times10^{-6})$ &$\cdots$ &$\leq2.25$ \\
$\mathcal{B}(\Lambda^+_c\to p\mu^+\mu^-)(\times10^{-8})$&$\leq7.7$&$\leq7.7$&$^{(2.8\pm0.4)\times10^{-4}}_{(3.77\pm2.28)\times10^{-6}}$ \cite{Faustov:2018dkn,Sirvanli:2016wnr}&$^{37\pm8}_{230\pm66}$ \cite{Faustov:2018dkn,Sirvanli:2016wnr} \\
$\mathcal{B}(\Xi^+_c\to \Sigma^+\mu^+\mu^-)(\times10^{-7})$&$\cdots$&$\leq1.25$ \\
$\mathcal{B}(\Xi^0_c\to\Lambda^0\mu^+\mu^-)(\times10^{-8})$&$\cdots$&$\leq8.42$ \\
$\mathcal{B}(\Xi^0_c\to\Sigma^0\mu^+\mu^-)(\times10^{-8})$&$\cdots$&$\leq2.15$ \\
$\mathcal{B}(\Xi^+_c\to p\mu^+\mu^-)(\times10^{-8}) $&$\cdots$&$\leq5.41$\\
$\mathcal{B}(\Xi^0_c\to n\mu^+\mu^-)(\times10^{-8}) $&$\cdots$&$\leq1.87$\\
\hline
\end{tabular}\label{Tab:BrTb32T8}}
\end{table}
From  Tab. \ref{Tab:SU3HATc3}, one can see  that singly Cabibbo suppressed $\Lambda^+_c\to p\ell^+\ell^-, \Xi^+_c\to \Sigma^+\ell^+\ell^-,\Xi^0_c\to\Lambda^0\ell^+\ell^-,\Xi^0_c\to\Sigma^0\ell^+\ell^-$ decays   receive both  the single-quark transition and  the $W$-exchange contributions, nevertheless, Cabibbo
allowed $\Lambda^+_c\to \Sigma^+\ell^+\ell^-,\Xi^0_c\to \Xi^0\ell^+\ell^-$ decays   and doubly Cabibbo suppressed $\Xi^+_c\to p\ell^+\ell^-,\Xi^0_c\to n\ell^+\ell^-$ decays only receive the $W$-exchange contributions.

In addition,
the contribution of $H(\bar{6})$ to the decay branching ratio is about 5.5 times larger than
one of $H(15)$ due to Wilson coefficient suppressed; for example, see Refs. \cite{Geng:2017esc,Geng:2018plk}.
If ignoring the Wilson coefficient suppressed $H(15)$ term contributions, there are only two parameters, $F_1$  and  $\widetilde{F}\equiv\widetilde{f}_1-\widetilde{f}_3$, in the decay amplitudes of  $T_{c3}\to T_{8}\ell^+\ell^-$.
The simplified results are listed in the last column of Tab. \ref{Tab:SU3HATc3}.
One can see that the  Cabibbo
allowed and doubly Cabibbo suppressed eight decay modes of $T_{c3}\to T_{8}\ell^+\ell^-$ are related by only one parameter $\widetilde{F}$;  nevertheless, the singly Cabibbo suppressed eight decays are related by two parameters $\widetilde{F}$ and $F_1$.  Moreover, there are amplitude relations $A(\Xi^+_c\to \Sigma^+ \ell^+\ell^- )=\sqrt{2}A(\Xi^0_c\to \Sigma^0 \ell^+\ell^-)$.

In these $T_{c3}\to T_{8}\ell^+\ell^-$ decays, only $\mathcal{B}(\Lambda^+_c\to pe^+e^-)$ and $\mathcal{B}(\Lambda^+_c\to p\mu^+\mu^-)$ are upper limited by experiment.  We list their experimental upper limits (exp. UL)  in the second column of Tab. \ref{Tab:BrTb32T8}.  Due to the lack of the experiment in $T_{c3}\to T_{8}\ell^+\ell^-$ decays and the  complex amplitude expressions included the $W$-exchange contributions,   we will only analyze the   $T_{c3}\to T_{8}\ell^+\ell^-$ decays  in the $S_1$ case. We assume that $W$-exchange contributions noted by $\widetilde{F}$ or the single-quark transition  contributions noted by $F_1$ play a dominant role in these decays. They  are separately discussed as follows.
\begin{itemize}
\item  Only considering  the $W$-exchange contributions by setting $F_1=0$,    all the $T_{c3}\to T_{8}\ell^+\ell^-$ decays  are related by one parameter $\widetilde{F}$ as shown in Tab. \ref{Tab:SU3HATc3}.    The upper limit predictions of $T_{c3}\to T_{8}e^+e^-$ and $T_{c3}\to T_{8}\mu^+\mu^-$ decays in the $S_1$ case are listed in the third  column of Tab. \ref{Tab:BrTb32T8}. One can see that  $\mathcal{B}(\Lambda^+_c\to p\mu^+\mu^-)$ gives an  effective constraint on these upper limit predictions of the branching ratios.

\item Only considering the single-quark transition  contributions by setting $\widetilde{F}=0$, as shown in Tab. \ref{Tab:SU3HATc3},  all eight  singly Cabibbo suppressed  $T_{c3}\to T_{8}e^+e^-$ and $T_{c3}\to T_{8}\mu^+\mu^-$ decays  are related by the parameter $F_1$.  Then the six upper limit predictions of $\mathcal{B}(\Lambda^+_c\to p e^+e^-)$, $\mathcal{B}(\Xi^+_c\to \Sigma^+ e^+e^-)$,   $\mathcal{B}(\Xi^0_c\to \Sigma^0 e^+e^-)$, $\mathcal{B}(\Lambda^+_c\to p \mu^+\mu^-)$, $\mathcal{B}(\Xi^+_c\to \Sigma^+ \mu^+\mu^-)$, and  $\mathcal{B}(\Xi^0_c\to \Sigma^0 \mu^+\mu^-)$ in the $S_1$ case have the same predictions  as the ones listed  in the third column of Tab. \ref{Tab:BrTb32T8}. Nevertheless, the predictions of  other two singly Cabibbo suppressed    $\mathcal{B}(\Xi^0_c\to \Lambda^0 e^+e^-)$  and $\mathcal{B}(\Xi^0_c\to \Lambda^0 \mu^+\mu^-)$  are different from the ones listed in  Tab. \ref{Tab:BrTb32T8}.  We obtain that $\mathcal{B}(\Xi^0_c\to \Lambda^0 e^+e^-)\leq9.66\times10^{-9}$  and $\mathcal{B}(\Xi^0_c\to \Lambda^0 \mu^+\mu^-)\leq9.36\times10^{-9}$. The previous other  predictions for $\mathcal{B}(\Lambda^+_c\to pe^+e^-)$ and $\mathcal{B}(\Lambda^+_c\to p\mu^+\mu^-)$, which considered the single-quark contributions with/without LD contributions,  are also listed in the last two columns of Tab. \ref{Tab:BrTb32T8} for comparing. The predicted upper limits of $\mathcal{B}(\Lambda^+_c\to p\ell^+\ell^-)$ are about 5 orders larger than the predictions with the SD  contributions, smaller than the LD contributing ones, which might mean that the $W$-exchange transitions cancel out with the  LD contributions.

\end{itemize}

\subsection{$T_{8}$ semileptonic weak decays}

 Similar to $T_{8}\to T'_{8}\gamma$ radiative decays,  $T_{8}\to T'_{8} \ell^+\ell^-$ semileptonic decays receive   the single-quark transition contributions and the $W$-exchange contributions  \cite{Bergstrom:1987wr}. The SD contributions come from the $Z^0$ and electromagnetic  penguin diagrams as well as the  $Z^0$ box  diagrams, and the LD contributions arise from  an intervirtual photon   in the $T_{8}\to T'_{8} \gamma^*$ processes.
 The LD contributions are much larger than the SD ones in the single-quark transition contributions in the $T_{8}\to T'_{8} \ell^+\ell^-$ decays.  So the LD contributions  and the W-exchange contributions might play the major roles in $T_{8}\to T'_{8} \ell^+\ell^-$ decays \cite{He:2005yn,Bergstrom:1987wr}.
The SU(3) flavor structure of the $s\to d$ Hamiltonian can be found in Ref. \cite{Wang:2019alu}.
The  SU(3) IRA  hadronic helicity amplitudes for $T_{8}\to T'_{8}\ell^+\ell^-$ decays via  $s\to d \ell^+\ell^-$ are
\begin{eqnarray}
H(T_{8} \to T'_8\ell^+\ell^-)^{L(R),s_p,s_k}_{VA,\lambda}&=&
   g_1(T_{8})^{[ij]n}T''(\bar{3})^k(T'_8)_{[ij]k}+g_2(T_{8})^{[ij]n}T''(3)^k(T_8)_{[ik]j}\nonumber\\
&+&g_3(T_{8})^{[in]j}T''(\bar{3})^k(T'_8)_{[ij]k}+g_4(T_{8})^{[in]j}T''(\bar{3})^k(T'_8)_{[ik]j}\nonumber\\
&+&g_5(T_{8})^{[in]j}T''(\bar{3})^k(T'_8)_{[jk]i}
  +\widetilde{g}_1(T_{8})^{[ij]n}(T'_8)_{[il]k}H(4)^{lk}_j \nonumber\\
&+&\widetilde{g}_2(T_{8})^{[in]j}(T'_8)_{[il]k}H(4)^{lk}_j
  +\widetilde{g}_3(T_{8})^{[jn]i}(T'_8)_{[il]k}H(4)^{lk}_j,\label{Eq:HT82T8p}
\end{eqnarray}
where  the model and scale independent parameters $g_i\equiv(g_i)^{L(R),s_p,s_k}_{VA,\lambda}$ and $\widetilde{g}_i\equiv(\widetilde{g}_i)^{L(R),s_p,s_k}_{VA,\lambda}$,  $T''(\bar{3})=(0,1,0)$ related to the transition operator $(\bar{d}s)$, and $H(4)^{lk}_j$ related to $(\bar{q}_lq^j)(\bar{q}_kq^n)$ operator with $n\equiv 3$ for $s$ quark  is symmetric in upper indices \cite{Wang:2019alu}.
The SU(3) IRA hadronic helicity amplitudes of $T_{8}\to T'_{8}\ell^+\ell^-$ weak decays are given in Tab. \ref{Tab:SU3HAT8},  in which
the information of the same CKM matrix elements  $V_{us}V^*_{ud}$ is not shown.
\begin{table}[t]
\renewcommand\arraystretch{1.3}
\tabcolsep 0.2in
\centering
\caption{The SU(3) IRA amplitudes of the  $T_{8}\to T'_{8}\ell^+\ell^- $ weak decays, $G_1\equiv g_1+g_2+g_3-g_5$, $G_2\equiv g_4+g_5$,   $\widetilde{G}_A\equiv \widetilde{g}_1-\widetilde{g}_3$, and  $\widetilde{G}_B\equiv \widetilde{g}_2+\widetilde{g}_3$.}\vspace{0.1cm}
{\footnotesize
\begin{tabular}{lcc}  \hline\hline
Decay modes~~~~~~~~~~~~& $A(T_{8}\to T'_{8}\ell^+\ell^-)$  \\\hline
$\Xi^-\to \Sigma^- \ell^+\ell^-  $&$G_1$\\
$\Xi^0\to \Lambda^0  \ell^+\ell^-  $&$(G_1+2G_2)/\sqrt{6}$\\
$\Xi^0\to \Sigma^0  \ell^+\ell^-  $&$(G_1+2\widetilde{G}_A)/\sqrt{2}$\\
$\Lambda^0\to n \ell^+\ell^-  $&$-\big[(G_1+2G_2)+(C_1+2\widetilde{G}_A)-(G_2-\widetilde{G}_B)\big]/\sqrt{6}$ \\
$\Sigma^0\to n \ell^+\ell^-  $&$-(G_2-\widetilde{G}_B)/\sqrt{2}$\\
$\Sigma^+\to p \ell^+\ell^-  $&$-(G_2+\widetilde{G}_B)$ \\\hline
\end{tabular}\label{Tab:SU3HAT8}}
\end{table}

\begin{table}[t]
\renewcommand\arraystretch{1.3}
\tabcolsep 0.2in
\centering
\caption{Branching ratios of $T_{8}\to T'_{8}\ell^+\ell^-$ decays  within $1\sigma$ theoretical error in the $S_1$ case. }\vspace{0.1cm}
{\footnotesize
\begin{tabular}{rcc}  \hline\hline
Decay modes~~~~~~~~~~~~& Experimental data \cite{PDG2020} &  Our predictions without $\widetilde{G}_{A,B}$ in $S_1$ \\\hline
$\mathcal{B}(\Xi^-\to \Sigma^-e^+e^-)(\times10^{-6})$&$\cdots$&$2.49^{+0.31}_{-0.29}$ \\
$\mathcal{B}(\Xi^0\to \Lambda^0 e^+e^-)(\times10^{-6})$&$7.6\pm0.6$&$7.6\pm0.6$\\
$\mathcal{B}(\Xi^0\to \Sigma^0 e^+e^-)(\times10^{-6})$&$\cdots$&$2.05\pm0.17$\\
$\mathcal{B}(\Lambda^0\to ne^+e^-)(\times10^{-5})$&$\cdots$&$2.06^{+0.25}_{-0.23}$ \\
$\mathcal{B}(\Sigma^0\to ne^+e^-)(\times10^{-17})$&$\cdots$&$7.61^{+6.59}_{-4.44}$ \\
$\mathcal{B}(\Sigma^+\to pe^+e^-)(\times10^{-7})$&$<70$&$1.60^{+1.14}_{-0.87}$ \\\hline
$\mathcal{B}(\Sigma^0\to n\mu^+\mu^-)(\times10^{-17})$&$\cdots$&$1.22^{+1.05}_{-0.71}$ \\
$\mathcal{B}(\Sigma^+\to p\mu^+\mu^-)(\times10^{-8})$&$2.4^{+1.7}_{-1.3}$&$2.40^{+1.70}_{-1.30}$ \\\hline
\end{tabular}\label{Tab:Tb32T8}}
\end{table}

There are four complex parameters $G_1,G_2,\widetilde{G}_A$ and $\widetilde{G}_B$ in  $T_{8}\to T'_{8}\ell^+\ell^- $ weak decays.
Since the initial baryon $\Xi^-$ does not contain $u$ quark and   the $W$-exchange contributions are canceled in $\Xi^0\to \Lambda^0  \ell^+\ell^-$ decays,
 the amplitudes of
$\Xi^-\to \Sigma^- \ell^+\ell^- $ and $\Xi^0\to \Lambda^0  \ell^+\ell^- $  listed in  Tab. \ref{Tab:SU3HAT8}  only contain coefficients $G_{1,2}$,  which means that the $W$-exchange transitions do not contribute to the $\Xi^-\to \Sigma^- \ell^+\ell^- $ and $\Xi^0\to \Lambda^0  \ell^+\ell^- $ decays. Therefore,  $\Xi^-\to \Sigma^- \ell^+\ell^- $ and $\Xi^0\to \Lambda^0  \ell^+\ell^- $ decays could be used to explore the LD contributions.
Other decay amplitudes contained both $G_{1,2}$ and $\widetilde{G}_{A,B}$ could proceed from the  LD contributions and the  W-exchange contributions.

Only two branching ratios  of $\Xi^0\to \Lambda^0 e^+e^-$ and $\Sigma^+\to p\mu^+\mu^-$ decays have been measured at present, which are listed in the second column of Tab. \ref{Tab:Tb32T8}.
We may  constrain $|G_1+2G_2|$ and $|G_2+\widetilde{G}_B|$ from the experimental data of $\mathcal{B}(\Xi^0\to \Lambda^0 e^+e^-)$ and $\mathcal{B}(\Sigma^+\to p\mu^+\mu^-)$, respectively, and we obtain that $\frac{|G_1+2G_2|}{|G_2+\widetilde{G}_B|}\approx12$.   $\mathcal{B}(\Sigma^+\to pe^+e^-)$ is obtained  by  using  the constrained $|G_2+\widetilde{G}_B|$, 
\begin{eqnarray}
\mathcal{B}(\Sigma^+\to pe^+e^-)=(1.60^{+1.14}_{-0.87})\times10^{-7},
\end{eqnarray}
which is  1 order smaller than its  experimental upper limits  $\mathcal{B}(\Sigma^+\to pe^+e^-)\leq7\times10^{-6}$ at the 90\% confidence level \cite{PDG2020}  and  is also smaller than  its SM predictions with the single-quark transition  LD contributions, $9.1\times10^{-6}\leq\mathcal{B}(\Sigma^+\to pe^+e^-)\leq10.1\times10^{-6}$   in Ref. \cite{He:2005yn}.

It is difficult to estimate which term gives the main contribution among  $G_1,G_2,\widetilde{G}_A$, and $\widetilde{G}_B$ now. Nevertheless, the  LD contributions noted by $G_{1,2}$ can not been entirely ignored via the experimental measurement of $\Xi^0\to \Lambda^0  \ell^+\ell^- $. So we will give the following discussions.
\begin{itemize}
\item If only considering the single-quark transition contributions, $i.e.$, $\widetilde{G}_A=\widetilde{G}_B=0$, one obtains $\left|\frac{G_1}{G2}+2\right|\approx12$; $i.e.$, $G_1\approx10G_2$ or $-14G_2$.  After  ignoring the small $G_2$ terms  in $G_1+2G_2$ and $2G_1+G_2$, one gets all branching ratios of  relevant $T_{8}\to T'_{8}\ell^+\ell^-$ weak decays  in  $S_1$ case,  which  are given in the last column of  Tab. \ref{Tab:Tb32T8}.

\item  If $|\widetilde{G}_B|\gg |G_2|$, $\frac{|G_1+2G_2|}{|G_2+\widetilde{G}_B|}\approx\frac{|G_1|}{|\widetilde{G}_B|}\approx12$, the predictions of $\mathcal{B}(\Xi^-\to \Sigma^-e^+e^-),~\mathcal{B}(\Xi^0\to \Lambda^0 e^+e^-), ~\mathcal{B}(\Sigma^0\to ne^+e^-),~\mathcal{B}(\Sigma^+\to pe^+e^-),~\mathcal{B}(\Sigma^0\to n\mu^+\mu^-)$, and $\mathcal{B}(\Sigma^+\to p\mu^+\mu^-)$ are the same as given in Tab. \ref{Tab:Tb32T8}.  Nevertheless,  $B\mathcal{}(\Xi^0\to \Sigma^0  \ell^+\ell^-)  $ and $\mathcal{B}(\Lambda^0\to n \ell^+\ell^-  )$ can not be predicted in this case.

\item In the case of  $\tilde{G}_B\approx G_2$, the predicted results are similar to above ones with  $|\widetilde{G}_B|\gg |G_2|$ except $\mathcal{B}(\Sigma^0\to n\ell^+\ell^-)\approx0$.

\item If $\widetilde{G}_B\approx -G_2$, $i.e.$, the contributions between  $G_2$ and  $\widetilde{G}_B$    are largely  canceled in $G_2+\widetilde{G}_B$ term,  the situations are complex,  which is beyond the scope of this paper.

\end{itemize}
 All predicted branching ratios except $\mathcal{B}(\Sigma^0\to ne^+e^-)$ and $\mathcal{B}(\Sigma^0\to n\mu^+\mu^-)$ in above first three cases are on the order of $\mathcal{O}(10^{-8}-10^{-5})$, some of them might be observed by BESIII and  Belle-II experiments  in the near future.
The measurement of $\mathcal{B}(\Xi^-\to \Sigma^-e^+e^-)$ and  $\mathcal{B}(\Xi^0\to \Sigma^0 e^+e^-)$ in the  future could further help us to understand the LD contributions and   the $W$-exchange contributions, respectively.

\section{SUMMARY}
Semileptonic baryon decays induced by  flavor changing neutral current  transitions play very important roles in testing the SM and  probing  new physics.
We  have studied the semileptonic decays of baryons with $\frac{1}{2}$ spin via the single-quark transitions  $b\to s/d\ell^+\ell^-$,    $c\to u\ell^+\ell^-$ and $s\to d\ell^+\ell^-$  as well as relevant  $W$-exchange transitions by using the  SU(3) flavor symmetry,  which is a powerful tool to test the  physics and to connect the physical quantities without knowing the underlying dynamics.
Our main results can be summarized as follows.

\begin{itemize}
\item{\bf $T_{b3}\to T_8 \ell^+\ell^-$  decays}:
Decay $\Lambda^0_b\to \Sigma^0\ell^+\ell^-$ is not allowed by  the  SU(3) irreducible representation approach, and  all other  21 decay amplitudes of the $T_{b3}\to T_8 \ell^+\ell^-$  decay modes via the $b\to s/d \ell^+\ell^-$  transitions  could  be  related by only one SU(3) flavor symmetry parameter, which could be constrained by the present experimental data of $\mathcal{B}(\Lambda^0_b\to \Lambda^0\mu^+\mu^-)$. Using the constrained   parameter,  we have predicted the not-yet-measured observables in the whole $q^2$ region and in different $q^2$ bins within the $S_1$ and $S_2$ cases.
The predicted branching ratios are on the order of  $\mathcal{O}(10^{-8}-10^{-6})$, many of them  are obtained for the first time,   and some of them  could be reached by the LHCb or Belle-II experiments.
The longitudinal polarization fractions and the
leptonic forward-backward asymmetries of  all $T_{b3}\to T_{8}\ell^+\ell^-$ decays are very similar to each other in certain $q^2$ bins  by the SU(3) flavor symmetry.  The predictions of  $\langle f_L\rangle(\Lambda^0_b\to \Lambda^0\mu^+\mu^-)_{[15,20]}$  and $\langle A^\ell_{FB}\rangle(\Lambda^0_b\to \Lambda^0\mu^+\mu^-)_{[15,20]}$   are agreeable with their experimental data within $1.5\sigma$   and  $1\sigma$ error ranges, respectively.

\item{\bf $T_{c3}\to T_8 \ell^+\ell^-$  decays}:
$T_{c3}\to T_8 \ell^+\ell^-$  decays are quite different from $T_{b3}\to T_8 \ell^+\ell^-$  decays, since the  former  may receive  both the  single-quark  $c\to u\ell^+\ell^-$  transition contributions and the $W$-exchange contributions.  After  ignoring the Wilson coefficient suppressed $H(15)$ terms, all decay amplitudes of  $T_{c3}\to T_{8}\ell^+\ell^-$ have been  related by  two SU(3) flavor symmetry parameters.  Using the 90\% experimental upper limit of  $\mathcal{B}(\Lambda^+_c\to p\mu^+\mu^-)$,   we have  obtained  the  upper limit predictions of the not-yet-measured $\mathcal{B}(T_{c3}\to T_{8}\ell^+\ell^-)$ by considering only one kind of  dominant contributions from   either  single-quark transition LD contributions or the $W$-exchange contributions.

\item{\bf $T_{8}\to T'_8 \ell^+\ell^-$  decays}:   Decays $T_{8}\to T'_8 \ell^+\ell^-$  are more complicated than both $T_{b3}\to T_8 \ell^+\ell^-$ and $T_{c3}\to T_8 \ell^+\ell^-$ ones, since the quarks
are antisymmetric in both the initial states $T_8$ and the final states $T'_8$. We have predicted  $\mathcal{B}(\Sigma^+\to pe^+e^-)$  in the $S_1$ case.  Moreover, we have analyzed the single-quark transition LD  contributions and  the $W$-exchange contributions, and found that $\mathcal{B}(\Xi^-\to \Sigma^-e^+e^-)$, $\mathcal{B}(\Xi^0\to \Sigma^0 e^+e^-)$, and  $\mathcal{B}(\Lambda^0\to ne^+e^-)$ are on the order of $\mathcal{O}(10^{-6}-10^{-5})$ in most cases except $\widetilde{G}_B\approx -G_2$.

\end{itemize}
According to our predictions,  many results in this work can be tested by the
experiments at BESIII,  LHCb, and Belle-II.
 And these results  can be used to test SU(3) flavor symmetry approach in $T_{b3,c3,8}\to T'_8\ell^+\ell^-$ by the future experiments.

\section*{ACKNOWLEDGEMENTS}
The work was supported by the National Natural Science Foundation of China (Contract No. 11675137).

\section*{References}

\end{document}